# Prediction of amino acid content in live black soldier fly larvae using near infrared spectroscopy

R.M. Zaalberg[1], L.B. Andersen[1], S.J. Noel[2], A.J. Buitenhuis[1], K. Jensen[2], G. Gebreyesus[1*]

[1]*Center for Quantitative Genetics and Genomics, C. F. Møllers Allé 3, 8000 Aarhus;*

[2]*Department of Animal and Veterinary Sciences, Blichers Allé 20, 8830 Tjele, Aarhus University, Denmark.*

[*]Corresponding author: grum.gebreyesus@qgg.au.dk

## Abstract

Black soldier fly (*Hermetia illucens*) larvae are gaining recognition as a sustainable farmed protein source for animal feed and human nutrition. Amino acid composition critically determines protein quality and overall nutritional value, necessitating rapid, cost-effective methods to ensure consistency in insect-based products. Non-destructive techniques to assess protein composition in live larvae are especially important for selective breeding purposes. This study establishes near-infrared (NIR) spectroscopy coupled with partial least squares (PLS) regression as a robust tool for predicting free amino acid (FAA) profiles in live black soldier fly (BSF) larvae. Larvae were reared on 17 diets with varying protein (2.3–89.8%) and sugar (8.5–96.0%) levels. A total of 204 samples of living larvae (111 - 337 larvae per sample) were scanned twice using a FOSS DS2500 spectrometer (400–2 500 nm) to capture spectral data, followed by freeze-drying and FAA extraction via methanol-based hydrolysis. We observed distinct FAA profiles, with alanine (6.50 ± 3.52 mg/g) and proline (5.61 ± 4.45 mg/g) as the most dominant components. Key NIR absorption bands at 1 670–1 786 nm and 2 300–2 366 nm was critical for predictions. PLS models optimized via repeated ten-fold cross-validation demonstrated high accuracy for individual FAAs (e.g., glutamate: $R^2$ = 0.88; asparagine: $R^2$ = 0.78; alanine: $R^2$ = 0.65) and the total free amino acid ($R^2$ = 0.64). These findings illustrate NIR's utility in BSF production and selective breeding, offering a rapid, scalable solution for real-time quality control as well as accurate phenotyping.

*Key words*: BSF larvae, nutritional composition, phenotyping, non-destructive, insect breeding.

**Implications**

Insect protein is a new sustainable food source for the growing world population. The larvae of the black soldier fly (*Hermetia illucens*) are exceptionally efficient at converting industrial waste products to valuable protein. There is a need to develop new methods to assess the nutritional composition of larvae, without having to kill the larvae. We show that it is possible to predict detailed amino acid content using infrared analysis of live larvae. Our study presents a promising method for the insect industry, especially for insect breeders who would like to select for larvae with a high-quality amino acid composition.

**Introduction**

The growing global demand for sustainable protein sources has led to increased interest in insects as alternative feed and food ingredients (van Huis et al., 2013). Among these, black soldier fly (*Hermetia illucens*) (**BSF**) larvae have gained particular attention due to their ability to efficiently convert organic waste into high-quality protein and fat (Barragán-Fonseca et al., 2017). Despite the increasing interest in commercial BSF production and its promising future, the industry is still in its infancy. There is room for improvement regarding the efficiency of insect production, for example through the development of new methods for monitoring the nutritional composition of insect products at different production stages.

Insect products are rich in protein, fat, and essential nutrients, making them valuable ingredients for animal feed. Maintaining consistent nutritional quality is crucial for the profitability and sustainability of insect farming (Lu et al., 2024). Among these, amino acid composition is a key determinant of insect protein quality, influencing digestibility, palatability, bioavailability, and the nutritional adequacy of feed formulations (Rumpold & Schlüter, 2013). BSF larvae provides a rich source of essential amino acids, including lysine and methionine, making it a promising alternative to conventional protein ingredients in livestock, pet, and aquaculture feed, where amino acid balance is crucial for optimal animal growth and health. Since the amino acid profile of BSF larvae is highly dependent on diet and developmental stage, routine monitoring is necessary to maintain consistent product quality (Spranghers et al., 2017). One common approach for quantifying amino acids involves the extraction of free amino acids (FAAs) and analysis via liquid chromatography-tandem mass spectrometry (LC-MS/MS) (Mak et al., 2019). Although this chemometric method offers sensitivity and specificity, it involves extensive sample preparation, including derivatization and careful handling to minimize matrix effects, which can make large-scale implementation more resource-intensive (Violi et al., 2020). Additionally, selective breeding presents a promising strategy for enhancing the nutritional quality of insect products. This approach requires the use of nondestructive methods to assess protein and fat content in live selection candidates (Ssemakula et al., 2024).

Near-infrared (**NIR**) spectroscopy has emerged as a promising non-destructive alternative for rapid, cost-effective evaluation of the biochemical composition of food and feed ingredients, including protein sources (Manley, 2014). By detecting molecular vibrations, NIR spectroscopy offers real-time nutritional assessment and has shown potential for predicting protein, lipid, and moisture content in insect-based feeds, particularly in BSF larvae (Alagappan et al., 2022; Hoffman et al., 2022; Cruz-Tirado et al., 2022; 2023a; 2023b; Kröncke et al., 2023). Recent studies have attempted to predict detailed fatty acid composition of larvae using NIR spectroscopy (Alagappan et al., 2025; Cruz-Tirado et al.,2025). To our knowledge, however, no study has yet applied NIR spectroscopy to quantify detailed FAA contents in live insects for food and feed. Notably, all previous investigations have primarily used dead samples, which limits the utility of these methods for selective breeding purposes.

In this study, we explore the feasibility of using NIR spectroscopy to predict FAA content in live BSF larvae reared on diets with varying protein and sugar levels. We employ partial least squares (**PLS**) regression modeling to evaluate the predictive accuracy of NIR-derived spectra for determining FAA concentrations. Successful implementation of this approach could provide a rapid, non-invasive tool for optimizing insect rearing strategies, improving feed formulations, and supporting selective breeding programs for enhanced nutritional quality of BSF products.

**Materials and Methods**

An overview of the experiment to obtain NIR spectra from living BSF larvae, and their FAA content is illustrated in Figure 1. BSF Larvae were reared on 17 diets with varying protein and carbohydrate content. Samples, each containing 111 - 337 individual larvae, were scanned with an NIR scanner and stored for chemical analysis between days 12–15 post-hatching. The experiment was repeated three times, yielding a total of 204 samples.

*Larvae rearing and dietary treatments*

The BSF larvae were collected at Enorm Biofactory A/S (Hedelundvej 15, 8762 Flemming, Denmark) at an age of 5-6 days and were then reared on 17 diets. The diets were based on chicken feed (Paco Start, Danish Agricultural Grocery Company, Fredericia, DK), to which casein or sucrose had been added at different proportions (Table 1). The diets were prepared the day before the larvae arrived. For each diet, four transparent plastic cups (70:95 mm H:Ø) were prepared with 60 grams of dry feed, 140 ml of water, and 1 gram of agar (Table 1). Afterwards, ~300 six-day old larvae were added to each cup. The cups were closed off with a lid with a 65 mm Ø hole covered with a fine mesh for ventilation. The cups were placed in a climate chamber, where the larvae were reared under controlled conditions (28°C, 70% RH) until harvesting. The larvae were harvested on four consecutive days: 12, 13, 14 and 15 days after egg hatching. On each harvesting day, for each of the 17 dietary treatments, one of the four cups was taken out of the climate chamber. After a cup was taken out, the larvae were rinsed with water and dried with paper towels. A randomly selected sample of live larvae was put in a quartz cup for the NIR-scanning. The live larvae were moved to a paper towel and placed back in the cup again for a second scanning (repacking). After the second scanning, the sample of larvae was weighed, and the number of larvae was determined upon visual inspection. Finally, the sample of scanned larvae was placed in a small plastic zip-lock bag, and moved to a freezer at a temperature of –20 °C. The larvae were freeze-dried, their dry weight was determined, and the samples were crushed for the chemical analysis.

### *NIR spectral acquisition*

A random selection of live larvae from a cup was scanned twice using a FOSS DS2500 NIR spectrometer. The small quartz cup was filled with as many larvae as possible to minimize the surface area without any larvae (pin hole gaps). The quartz cup was gently closed off with a metal lid and placed in the scanner. After the first scanning, the quartz cup was repacked by removing the larvae and putting them back in, and the larvae were scanned a second time. Upon scanning, the quartz cup contained between 111 and 337 larvae, depending on the dietary treatment. Data recording was done every 0.5 nm from 400 to 2 500 nm.

The duplicate packing samples were scanned twice, and the resulting spectra were averaged before preprocessing to enhance signal quality and optimize model performance. Multiple spectral preprocessing techniques were applied sequentially to minimize noise and scattering effects. First, Savitzky-Golay smoothing (first derivative, polynomial order 2, window size 21) was used to reduce noise while preserving spectral features (Savitzky & Golay, 1964). Standard normal variate transformation was then applied to correct for baseline shifts and scattering variations (Barnes et al., 1989). Finally, multiplicative scatter correction was implemented to normalize spectral differences due to physical variations in sample presentation. Different combinations of spectral preprocessing were tested for each component and the best performing models were selected.

*Freeze-drying and chemical extraction of free amino acids*

After NIR scanning, samples were freeze-dried and then homogenized using an Emerio MC126196 blender for 1–2 minutes, followed by manual grinding with a mortar and pestle. Approximately 100 mg of freeze-dried material was transferred into 1.5 mL screw-cap tubes, followed by the addition of 1 mL of 60% methanol. Two metal beads were added to each tube, and the samples were homogenized in a Retsch MM 400 mixer mill at 20 Hz for 10 minutes. The homogenates were centrifuged at 17,000 × g for 10 minutes, and the supernatant was collected. This supernatant was diluted 5-fold in deionized water and filtered through 0.22 μm pore-size filter plates to remove particulates.

*LC-MS/MS analysis of amino acids*

Liquid chromatography coupled with LC-MS/MS analysis was performed at the DynaMo MS-Analytics Facility at the University of Copenhagen (Denmark). Prior to analysis, the samples were further diluted 10-fold in deionized water and spiked with internal standards, including a 13C,15N-labeled amino acid mix (Merck Sigma-Aldrich Cell Free Amino Acid Mixture - 13C,15N 767964-1EA). Chromatography was performed on a 1290 Infinity II UHPLC system (Agilent Technologies). Separation was achieved on a Zorbax Eclipse XDB-C18 column (150 x 3.0 mm, 1.8 μm, Agilent Technologies). Formic acid (0.05 %, v/v) in water and acetonitrile (supplied with 0.05 % formic acid, v/v) were employed as mobile phases A and B respectively. The

elution profile r was: 0.00-0.75 min, 3 % B; 0.75-4.10 min, 3-65 % B; 4.10-4.20 min, 65-100 % B; 4.20-4.95 min, 98 % B; 4.95-5.00 min, 100-3 % B; and 5.0-6.0 min 3 % B. The mobile phase flow rate was 400 µL/min. The column temperature was maintained at 40 °C. The liquid chromatography was coupled to an Ultivo Triplequadrupole mass spectrometer (Agilent Technologies) equipped with a Jetstream electrospray ion source operated in positive ion mode. The instrument parameters were optimized by infusion experiments with pure standards. The ion spray voltage was set to +2500 V in positive ion mode. Dry gas temperature was set to 325 °C and dry gas flow to 11 L/min. Sheath gas temperature was set to 350 °C and sheath gas flow to 12 L/min. Nebulizing gas was set to 40 psi. Nitrogen was used as dry gas, nebulizing gas and collision gas. Multiple reaction monitoring was used to monitor precursor ion → fragment ion transitions. Multiple reaction monitoring transitions and other parameters such as fragmentor voltage and collision energies were optimized using reference standards. Both Q1 and Q3 quadrupoles were maintained at unit resolution. Multiple reaction monitoring (MRM) transitions, fragmentor voltage and collision energies are detailed in Supplementary File 1. Mass Hunter Quantitation Analysis for QQQ software (Version 10, Agilent Technologies) was used for data processing. Quantification of the individual amino acids was performed from the $^{13}C$, $^{15}N$-labeled amino acids used as internal standards.

*PLS regression model development*

NIR spectral data were correlated with FAA concentrations determined via LC-MS/MS to develop predictive models using PLS regression implemented in the "pls" package, version 2.8-5 (Liland et al., 2024), in R (R Core Team, 2024). The dataset was split into a 75% training set and a 25% independent test set using stratified partitioning to ensure representation across dietary treatments. Within the training set, a ten-fold repeated (x2) cross-validation (**CV**) strategy was implemented to optimize model complexity by selecting the number of components that maximized prediction accuracy while minimizing overfitting. To enhance robustness, model tuning was performed using a randomized grid search approach, selecting the optimal number of PLS components based on the R² and lowest root RMSE. The final PLS regression model was evaluated on the test set using key performance metrics, including the coefficient of determination in cross validation ($R^2$), standard error in cross validation (SECV), standard error of prediction (SEP), the residual predictive deviation (RPD = SD/SECV), bias and slope following (Williams et al., 2017).

To identify the spectral regions most influential for each prediction, a Variable Importance in Projection (**VIP**) function was implemented using the predictor loadings, scores, and Y-loadings extracted from the PLS regression model implemented as described above. VIP scores were then computed for each wavelength by weighting each predictor's contribution to the PLS model according to the sum of squared PLS loadings, scaled by the proportion of variance explained in the response variable. This ensures that wavelengths with a higher influence on the model receive higher VIP scores, allowing identification of key spectral regions. Additionally, residual plots were generated from test set predictions to assess model performance. All analyses were conducted in R using the pls package. Plots of VIP scores (Supplementary File 2) and residuals (Supplementary File 3) are provided as supplementary materials.

## Results

### *Free amino acid profiles in BSF larvae*

Table 2 presents the descriptive statistics of FAA concentrations (expressed in mg/g of sample) measured by LC–MS/MS for black soldier fly larvae reared on 17 different dietary treatments. Notable differences in both mean concentrations and variability (as indicated by the coefficient of variation, CV) were observed among the FAAs. For example, Alanine was the most abundant FAA, with a mean concentration of 6.50 mg/g (± 3.52) and a CV of 54.16%, while Asparagine was present at a much lower mean level (0.06 mg/g ± 0.08) yet exhibited a high CV of 124.7%. In contrast, essential amino acids such as lysine and leucine had moderate mean concentrations of 1.34 and 1.25 mg/g (CV of 61.52% and 56.08%, respectively). The total FAA (**tFAA**) had a mean concentration of 34.95 mg/g (± 17.29) with a CV of 49.46%, indicating substantial variability across diets.

### *PLS regression model performance*

Table 2 summarizes the predictive performance of the various FAA traits, which varied considerably among amino acids. An analysis of the relationship between mean FAA concentration and model performance ($R^2$) revealed a general trend where traits with higher concentrations tended to exhibit stronger predictive accuracy (Figure 2). For instance, glutamate and alanine, which had relatively high mean concentrations (2.41 mg/g and 6.50 mg/g, respectively), also demonstrated moderate to high $R^2$ values (0.88 and 0.65, respectively), indicating that the PLS regression model effectively captured their spectral variations. However, certain exceptions were observed; for example, proline, despite its relatively high concentration (5.61 mg/g), exhibited only moderate predictability (R² = 0.54), suggesting potential spectral complexities or interference.

A closer inspection of model residuals and additional performance metrics also revealed instances of overfitting and underfitting for certain FAAs. Glutamine, for example, exhibited a high ratio of performance to deviation (RPD = 35.12) but a relatively low R² (0.32) and a high standard error of cross-validation (SECV = 0.17), suggesting potential overfitting. This is further evidenced by its poor generalization, as indicated by a significant bias (0.28) and a low slope (0.27), implying that predictions systematically deviated from actual values.

Conversely, asparagine showed a relatively high R² (0.78) but a low concentration (0.06 mg/g) with a high coefficient of variation (124.7%). While this suggests that the model effectively captured spectral trends associated with asparagine, its low relative concentration raises concerns about whether the prediction is driven by true chemical signals or potential spectral artifacts. Similarly, proline, despite its relatively high concentration (5.61 mg/g), exhibited moderate predictability (R² = 0.54) and a high SECV (0.13), which may indicate spectral interference or overlapping absorbance patterns with other compounds.

Overall, while some FAAs exhibited signs of overfitting or lower predictive performance, the model successfully captured spectral variations for most FAAs. The strong predictive power observed for key FAAs highlights the potential of NIR spectroscopy for rapid FAA quantification in live BSF larvae, with opportunities for further refinement to enhance predictions for more challenging traits.

***Variable importance and robustness***

Our VIP analyses performed on each PLS regression model identified consecutive wavelength ranges where the VIP score exceeded a threshold (≥1.0) for each FAA. While precise intervals varied for each FAA, several wavelength ranges emerged repeatedly across multiple traits, often with VIP peaks exceeding the threshold. Table 3 presents an overview of the five most prominent intervals, spanning from the visible to near-infrared regions, and indicates the number FAAs with elevated VIP scores within these intervals, as well as the maximum VIP score observed.

In the visible range (405–450 nm), multiple FAAs (including alanine, aspartate, lysine, valine, phenylalanine, and tFAA) displayed moderate to high VIP values, with lysine reaching a maximum VIP of 3.04. A mid-visible region near 550–580 nm was also identified in several models (alanine, lysine, valine), albeit with somewhat lower maxima (up to about 2.22). However, the two intervals consistently yielding the highest VIP peaks, often above 3.0, were in the near-infrared range: approximately 1 670–1 786 nm and 2 300–2 366 nm. For instance, tryptophan reached a VIP of ~3.51 in the 1 670–1 786 nm interval, whereas Aspartate exhibited ~3.70 in the 2 300–2 366 nm region.

## Discussion

### *Nutritional profile of BSF larvae*

Our study represents the first comprehensive analysis of the FAA profile in BSF larvae, quantifying 19 individual amino acids using high-resolution LC-MS/MS. In contrast to previous investigations (e.g., De Marco et al., 2015; Spranghers et al., 2017; Barroso et al., 2017), which predominantly focused on the total amino acid composition of processed BSF larvae meals using conventional hydrolysis-based methods, our approach directly captures the FAA fraction in living larvae. This distinction is crucial, as hydrolysis releases both free and protein-bound amino acids while potentially introducing degradation artifacts (Violi et al., 2020). By contrast, our LC-MS/MS method quantifies only the non-bound FAAs, providing a dynamic and physiologically relevant snapshot of the larvae's metabolic state.

### *Spectral prediction of FAA*

Our study presents the first successful prediction of FAA content from NIR spectra of live BSF larvae, marking a novel advancement in insect nutritional analysis, particularly for selective breeding applications. The predictive performance of our PLS regression models was to some extent linked to the concentration and variability of individual FAAs. Higher concentration amino acids, such as alanine (6.50 mg/g) and glutamate (2.41 mg/g), yielded robust models with relatively high coefficients of determination ($R^2$ = 0.65 and 0.88, respectively) and lower prediction errors. In contrast, FAAs with very low concentrations or high relative variability, such as asparagine (CV = 124.7%), were more challenging to predict accurately. This relationship between concentration and model performance aligns with findings in other food matrices, where homogenized samples with higher analyte levels, such as milk amino acids (McDermott et al., 2016) and feedstuff fatty acid analyses (Fontaine et al., 2001), tend to produce more reliable NIR predictions.

However, our findings also highlight the importance of evaluating multiple performance metrics beyond $R^2$ alone, as a high coefficient of determination does not necessarily guarantee robust model generalization. For example, glutamine exhibited an extremely high ratio of performance to deviation (RPD = 35.12), but its relatively low $R^2$ (0.32) and low slope (0.27) suggest overfitting, meaning that while the model appeared to perform well within the training set, its ability to generalize to new samples was poor. Conversely, proline, despite its relatively high concentration (5.61 mg/g), showed only moderate predictability ($R^2 = 0.54$) and a high standard error of cross-validation (SECV = 0.13), indicating spectral complexities or overlapping absorbance patterns that limited accurate FAA quantification. These cases underscore the need for a multifaceted evaluation of model performance to distinguish genuinely strong predictions from those that may be misleading due to over- or underfitting.

While our work is novel in its focus on FAA prediction from live BSF larvae, research in related areas is rapidly advancing. Recent studies have reported promising attempts to predict detailed fatty acid composition from live BSF larvae using NIR spectroscopy (Alagappan et al., 2025) and hyperspectral imaging (Cruz-Tirado et al., 2025). These findings reinforce the growing potential of non-destructive spectral methods to assess complex nutritional profiles in intact insect samples. Larger datasets, advanced chemometric techniques, or hybrid modeling approaches integrating spectral and biological data could further refine prediction accuracy. Collectively, our results and those reported in the broader literature demonstrate that, although predicting the detailed nutritional composition of live larvae presents challenges, NIR spectroscopy remains a promising and rapidly evolving tool for non-destructive nutritional assessment in both traditional food systems and emerging insect production.

*Variable importance and robustness*

Our VIP analysis revealed that key NIR-regions, specifically between 1 670 and 1 786 nm and between 2 300 and 2 366 nm, consistently exhibited high VIP scores across a wide range of FAAs in BSF larvae. These regions likely reflect strong overtone and combination bands of –CH, –NH, and –OH bonds, which are critical for the chemical identification of amino acids. Notably, FAA models with robust PLS regression model performance (e.g., glutamate and asparagine) showed pronounced VIP peaks within these intervals, suggesting that these spectral regions capture fundamental chemical information necessary for accurate prediction. Similar findings have been reported in studies on BSF larvae. For example, Cruz-Tirado et al. (2025) demonstrated that, for the prediction of fatty acids in dried BSF larvae, the optimal PLS loadings for stearic acid were observed around 4 464 cm$^{-1}$, corresponding to approximately 2 240 nm, overlapping with our identified region near 2 300–2 366 nm. Furthermore, Alagappan et al. (2025) described that their spectrometer operated over a spectral range of 11,550 to 3,950 cm$^{-1}$ (corresponding to approximately 870 to 2530 nm), a range that encompasses our critical intervals. Although explicit wavelength intervals were not detailed in that study, the high performance of their calibration models ($R^2$ up to 0.78) implies that the most informative spectral bands likely fall within the near-infrared regions we identified. Together, these studies support the conclusion that targeting the 1 670–1 786 nm and 2 300–2 366 nm intervals may enhance model robustness and improve the non-destructive prediction of detailed nutritional profiles, including FAAs, in BSF larvae, offering a promising avenue for rapid quality control and selective breeding in insect production.

Despite promising results in the very first attempt of predicting detailed amino acid content in live BSF larvae in this study, several limitations warrant further investigation. First, we analyzed groups of larvae under the implicit assumption of homogeneity, potentially overlooking intra-group variability. This issue is compounded by the lack of controlled genetic structures, as larvae were not grouped by pedigree, which may have increased intra-group variability while minimizing differences between groups. Future studies should consider analyzing individual larvae and controlling the genetic background. Additionally evaluating alternative chemometric techniques and prediction models, including deep learning, could further enhance predictive accuracy.

**Conclusion**

This study presents a pioneering effort at the rapid prediction of free amino acids in live BSF larvae using NIR spectroscopy combined with PLS regression. By successfully correlating NIR spectral data with LC-MS/MS-quantified FAA concentrations, our models provide a non-destructive, rapid, and reliable method for assessing the nutritional quality of BSF larvae. The ability to quantify FAA composition in live larvae without the use of chemicals makes this approach not only efficient but also environmentally friendly, reinforcing its value in sustainable insect farming systems. The non-destructive nature of this method is particularly advantageous for selective breeding, enabling phenotyping of nutritional profiles in live selection candidates. Furthermore, the substantial predictive accuracy for some of the major amino acids across diverse dietary treatments underscores the robustness of NIR spectroscopy for real-time quality control and process optimization in BSF production. While this study demonstrates the feasibility of FAA prediction from live BSF larvae, future research should focus on further refining these models by incorporating larger datasets, accounting for environmental variability, and expanding their application to additional nutritional markers such as minerals. Advancements in machine learning integration may also enhance predictive performance, ultimately strengthening the role of NIR spectroscopy as a core technology for sustainable and scalable insect protein production.

**Ethics approval**

Not applicable.

**Data and model availability statement**

The data presented in this study will be made accessible in a public repository.

**Declaration of generative AI and AI-assisted technologies in the writing process**

The authors did not use any artificial intelligence assisted technologies in the writing process.

**Author ORCIDs**

R.M. Zaalberg: 0000-0002-2609-3458

L.B. Andersen: Not available

S. J. Noel: 0000-0003-4529-9598

A. J. Buitenhuis: 0000-0002-4953-3081

K. Jensen: 0000-0003-0261-3831

G. Gebreyesus: 0000-0003-4757-3060

**Declaration of interest**

Not applicable.

**Acknowledgements**

We would like to thank Enorm Biofactory A/S for providing the larvae. We would also like to acknowledge the DynaMo MS-Analytics Facility of the University of Copenhagen where LC-MS/MS analysis was performed.

**Financial support**

This study was part of the LaserLarvae project, funded by AUFF NOVA, Grant No: AUFF-E-2023-9-3

**Tables**

Table 1 – Nutritional composition of 17 diets used for the black soldier fly larvae. All diets were prepared by mixing 60 g of dry feed, 140 ml of water, and 1 g of agar.

| Diet | Components of dry feed (%) | | | Nutritional composition dry feed (%) | | | | |
|---|---|---|---|---|---|---|---|---|
| | Chicken feed | Casein | Sucrose | Protein | Carbs | Fat | Crude fibre | Crude ash |
| 1 | 12.5 | 87.5 | 0 | 89.8 | 8.5 | 0.5 | 0.6 | 0.7 |
| 2 | 25 | 75 | 0 | 79.6 | 17.0 | 0.9 | 1.2 | 1.3 |
| 3 | 33 | 67 | 0 | 73.1 | 22.4 | 1.2 | 1.6 | 1.7 |
| 4 | 50 | 50 | 0 | 59.3 | 34.0 | 1.8 | 2.4 | 2.7 |
| 5 | 67 | 33 | 0 | 45.4 | 45.5 | 2.4 | 3.1 | 3.6 |
| 6 | 75 | 25 | 0 | 38.9 | 50.9 | 2.7 | 3.5 | 4.0 |
| 7 | 87.5 | 12.5 | 0 | 28.7 | 59.4 | 3.2 | 4.1 | 4.6 |
| 8 | 93.75 | 6.25 | 0 | 23.6 | 63.7 | 3.4 | 4.4 | 5.0 |
| 9 | 100 | 0 | 0 | 18.5 | 67.9 | 3.6 | 4.7 | 5.3 |
| 10 | 93.75 | 0 | 6.25 | 17.3 | 69.9 | 3.4 | 4.4 | 5.0 |
| 11 | 87.5 | 0 | 12.5 | 16.2 | 71.9 | 3.2 | 4.1 | 4.6 |
| 12 | 75 | 0 | 25 | 13.9 | 75.9 | 2.7 | 3.5 | 4.0 |
| 13 | 67 | 0 | 33 | 12.4 | 78.5 | 2.4 | 3.1 | 3.6 |
| 14 | 50 | 0 | 50 | 9.3 | 84.0 | 1.8 | 2.4 | 2.7 |
| 15 | 33 | 0 | 67 | 6.1 | 89.4 | 1.2 | 1.6 | 1.7 |
| 16 | 25 | 0 | 75 | 4.6 | 92.0 | 0.9 | 1.2 | 1.3 |
| 17 | 12.5 | 0 | 87.5 | 2.3 | 96.0 | 0.5 | 0.6 | 0.7 |

Table 2. Average concentrations (mg/g of sample) of key free amino acids along with the predictive performance of partial least squares regression across the free amino acids and evaluated with different metrics.

| FAA[1] | Mean (± SD) | Range | CV (%) | R2 | SECV | SEP | RPD | Bias | Slope |
|---|---|---|---|---|---|---|---|---|---|
| Alanine | 6.50 (3.52) | 0.16-17.69 | 54.16 | 0.65 | 0.13 | 1.94 | 25.32 | -0.23 | 0.63 |
| Arginine | 1.09 (0.85) | 0.02-3.93 | 78.3 | 0.57 | 0.06 | 0.53 | 13.48 | 0.07 | 0.61 |
| Asparagine | 0.06 (0.08) | 0.001-0.35 | 124.7 | 0.78 | 0.02 | 0.03 | 3.75 | 0.00 | 0.78 |
| Aspartate | 0.33 (0.22) | 0.04-1.90 | 67.21 | 0.05 | 0.03 | 0.26 | 8.95 | -0.01 | 0.08 |
| Glutamate | 2.41 (1.66) | 0.14-8.18 | 68.78 | 0.88 | 0.08 | 0.69 | 22.96 | -0.03 | 0.74 |
| Glutamine | 5.89 (5.46) | 0.17-27.85 | 92.57 | 0.32 | 0.17 | 4.92 | 35.12 | 0.28 | 0.27 |
| Glycine | 1.16 (0.74) | 0.05-4.85 | 63.37 | 0.53 | 0.06 | 0.54 | 13.07 | 0.03 | 0.47 |
| Histidine | 2.32 (2.46) | 0.12-12.25 | 105.95 | 0.67 | 0.09 | 1.29 | 23.9 | 0.11 | 0.83 |
| Isoleucine | 0.96 (0.47) | 0.10-2.57 | 48.82 | 0.33 | 0.05 | 0.35 | 8.35 | -0.05 | 0.46 |
| Leucine | 1.25 (0.70) | 0.10-4.13 | 56.08 | 0.56 | 0.06 | 0.46 | 11.53 | 0.04 | 0.50 |
| Lysine | 1.34 (0.82) | 0.14-4.69 | 61.52 | 0.45 | 0.06 | 0.70 | 15.89 | -0.05 | 0.47 |
| Methionine | 0.20 (0.15) | 0.01-0.82 | 73.18 | 0.41 | 0.03 | 0.10 | 4.04 | 0.00 | 0.60 |
| Phenylalanine | 0.64 (0.29) | 0.12-1.60 | 45.51 | 0.25 | 0.04 | 0.28 | 8.13 | -0.02 | 0.30 |
| Proline | 5.61 (4.45) | 0.14-23.33 | 79.26 | 0.54 | 0.13 | 2.85 | 31.58 | -0.39 | 0.66 |
| Serine | 0.95 (0.63) | 0.07-5.41 | 66.25 | 0.17 | 0.05 | 0.79 | 17.49 | 0.09 | 0.17 |
| Threonine | 0.91 (0.35) | 0.2-1.98 | 38.89 | 0.32 | 0.05 | 0.28 | 6.71 | 0.00 | 0.26 |
| Tryptophane | 0.63 (0.26) | 0.12-1.26 | 40.95 | 0.44 | 0.03 | 0.19 | 8.69 | -0.01 | 0.42 |
| Tyrosine | 1.10 (0.81) | 0.04-3.19 | 73.16 | 0.32 | 0.06 | 0.7 | 13.86 | -0.03 | 0.44 |
| Valine | 1.48 (0.72) | 0.12-4.01 | 48.55 | 0.37 | 0.06 | 0.69 | 14.31 | 0.09 | 0.28 |
| tFAA | 34.95 (17.29) | 4.16-85.3 | 49.46 | 0.64 | 0.26 | 11.22 | 72.07 | 1.07 | 0.59 |

[1]Free amino acid expressed in units of mg/g of sample

tFAA = Total free amino acid as sum of values for all extracted FAAs in mg/g of sample; SD = Standard deviation; CV = Coefficient of variation (%); R² = Coefficient of determination; SECV = Standard error of cross-validation; SEP = Standard error of prediction; RPD = Ratio of performance deviation; Bias = Mean difference between predicted and observed values; Slope = Regression slope of predicted versus observed values.

Table 3. Summary of spectral intervals with high variable importance in projection (VIP) scores exceeding the threshold (>1.0) across multiple prediction models for predicting free amino acids (FAA) in the black soldier fly larvae.

| Spectral region (range in nm) | Number of FAAs with elevated VIP score | Max score (Value, FAA) |
|---|---|---|
| 405–450 | 12 | 2.22 (Lysine) |
| 550–580 | 7 | 3.04 (Lysine) |
| 1 204–1 220.5 | 17 | 1.84 (Aspartate) |
| 1 670–1 786 | 19 | 3.51 (Tryptophane) |
| 2 300–2 366 | 19 | 3.70 (Aspartate) |

**List of figures**

Figure 1. Overview of the experiment to obtain near infrared (NIR) spectra free amino acid (FAA) composition of black soldier fly larvae.

Figure 2. Relationship between mean FAA concentration in black soldier fly larvae and PLS regression model performance ($R^2$). *Figure 2 illustrates the relationship between the mean concentration of free amino acids (FAAs) and the predictive performance of the Partial Least Squares (PLS) regression model, as measured by $R^2$. Each point represents a different FAA, with unique shapes distinguishing individual amino acids. The general trend suggests that higher FAA concentrations are associated with improved model performance, though exceptions indicate potential spectral complexities influencing prediction accuracy.*

**List of supplementary files**

Supplementary File1: Multiple reaction monitoring (MRM) transitions, fragmentor voltage and collision energies used for amino acid analysis.

Supplementary File 2: Variable Importance in Projection (VIP) scores for free amino acids

This file contains VIP score plots for each free amino acid (FAA) trait analyzed in the study. The VIP scores were derived from Partial Least Squares (PLS) regression models, indicating the relative importance of different spectral wavelengths in FAA prediction. Each plot displays VIP scores across the full NIR spectral range (405–2494.5 nm), with a red dashed line at VIP = 1.0 as a threshold for identifying the most influential wavelengths. These plots help visualize which spectral regions contribute most to model predictions, aiding in the interpretation of spectral patterns linked to FAA composition in live BSF larvae.

Supplementary File 3: Residual plots for free amino acid predictions

This file presents residual plots for each FAA trait, illustrating the relationship between predicted and actual values from the PLS regression models. Residuals (the difference between observed and predicted values) are plotted against predicted concentrations to assess model accuracy and detect potential biases or trends. A red dashed line at zero residual value is included to indicate ideal prediction performance. These plots provide additional insights into model fit, highlighting instances of systematic errors, overfitting, or underfitting in specific FAA predictions.

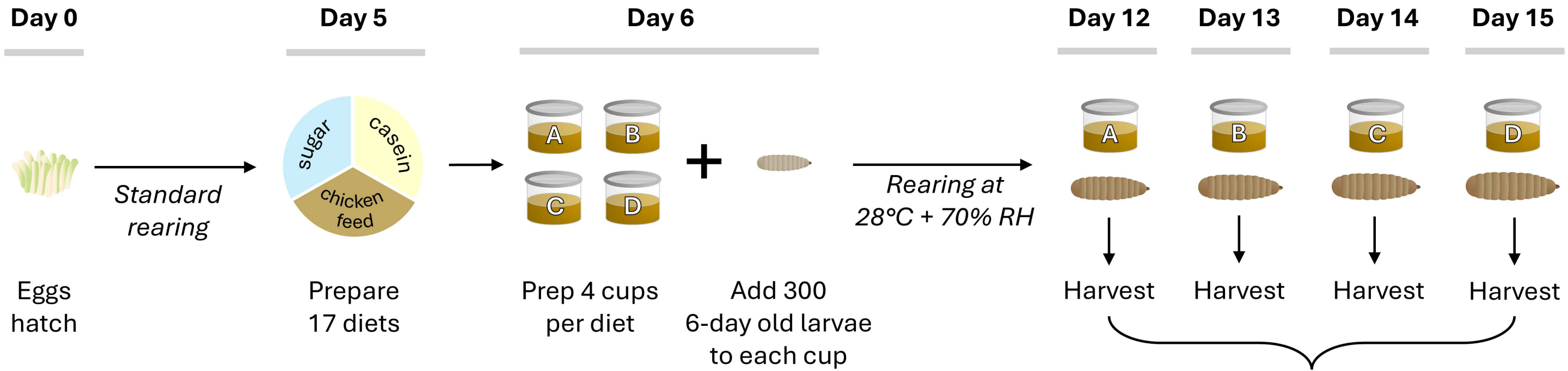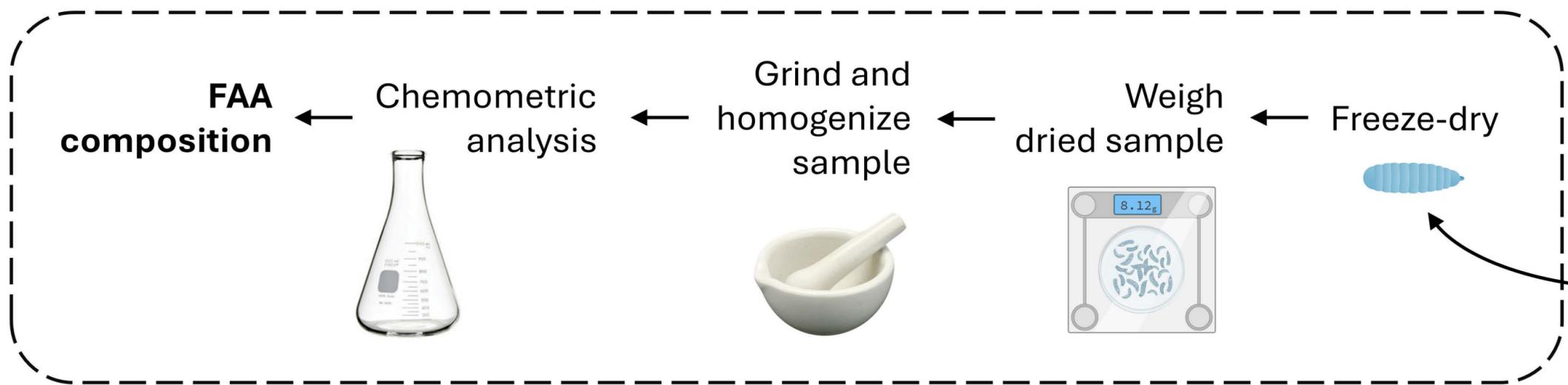

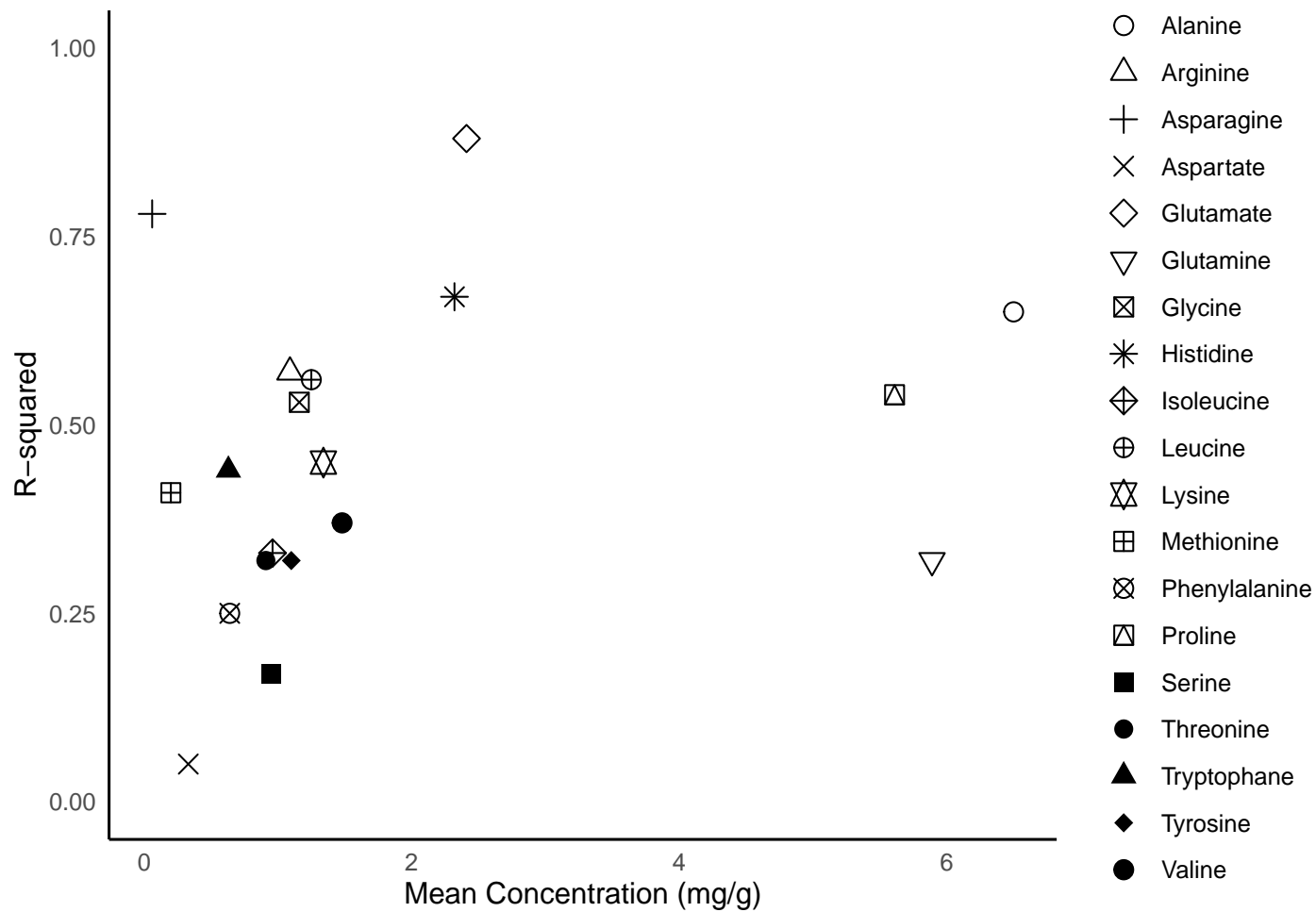

**Supplementary File1: Multiple reaction monitoring (MRM) transitions, fragmentor voltage and collision energies used for amino acid analysis**

| Compound name | Precursor ion (m/z) | Product ion (m/z) | Fragmentor (V) | CE (V) |
|---|---|---|---|---|
| 13C-15N-Ala | 94.1 | 47.1 | 63 | 8 |
| 13C-15N-Arg | 185.1 | 75.1 | 63 | 28 |
| 13C-15N-Asn | 139.1 | 77.1 | 63 | 12 |
| 13C-15N-Asp | 139.1 | 77.1 | 63 | 12 |
| 13C-15N-Cys | 126 | 79.2 | 63 | 8 |
| 13C-15N-Gln | 154.1 | 136 | 63 | 4 |
| 13C-15N-Glu | 154.1 | 107.1 | 63 | 8 |
| 13C-15N-Gly | 79 | 50.3 | 63 | 4 |
| 13C-15N-Gly | 79 | 32.5 | 63 | 4 |
| 13C-15N-His | 165.2 | 118.1 | 63 | 12 |
| 13C-15N-Ile | 139.2 | 92.1 | 63 | 8 |
| 13C-15N-Leu | 139.2 | 92.1 | 63 | 8 |
| 13C-15N-Lys | 155.1 | 90.1 | 63 | 16 |
| 13C-15N-Met | 156.1 | 109.1 | 63 | 8 |
| 13C-15N-Phe | 176.2 | 129.2 | 63 | 12 |
| 13C-15N-Pro | 122.1 | 75 | 63 | 16 |
| 13C-15N-Ser | 110 | 63.1 | 63 | 8 |
| 13C-15N-Thr | 125.1 | 78.2 | 63 | 8 |
| 13C-15N-Trp | 218.2 | 200.1 | 63 | 4 |
| 13C-15N-Trp | 218.2 | 156.2 | 63 | 8 |
| 13C-15N-Trp | 218.2 | 127.1 | 63 | 28 |
| 13C-15N-Tyr | 192.1 | 145.2 | 63 | 12 |
| 13C-15N-Val | 124.1 | 77.2 | 63 | 8 |
| Alanine | 90.1 | 44.1 | 63 | 8 |
| Arginine | 175.1 | 70.1 | 63 | 28 |
| Arginine | 175.1 | 60.1 | 63 | 12 |
| Asparagine | 133.1 | 74.1 | 63 | 12 |
| Aspartate | 134.4 | 74.1 | 63 | 12 |
| Glutamate | 148.1 | 102.1 | 63 | 8 |
| Glutamate | 148.1 | 84 | 63 | 12 |
| Glutamate | 148.1 | 56 | 63 | 32 |
| Glutamine | 147.1 | 130 | 63 | 4 |
| Glycine | 76 | 48.3 | 63 | 4 |
| Glycine | 76 | 30.5 | 63 | 4 |
| Histidine | 156.2 | 110.1 | 63 | 12 |
| Histidine | 156.2 | 93 | 63 | 24 |
| Isoleucine | 132.2 | 86.1 | 63 | 8 |
| Leucine | 132.2 | 86.1 | 63 | 8 |
| Lysine | 147.1 | 84.1 | 63 | 16 |
| Methionine | 150.2 | 104.1 | 63 | 8 |
| Methionine | 150.2 | 61.1 | 63 | 20 |
| Methionine | 150.2 | 56.1 | 63 | 12 |
| Phenylalanine | 166.2 | 120.2 | 63 | 12 |
| Phenylalanine | 166.2 | 103.1 | 63 | 28 |
| Proline | 116.1 | 70 | 63 | 16 |
| Proline | 116.1 | 43 | 63 | 0 |
| Serine | 106 | 60.1 | 63 | 8 |
| Serine | 106 | 42.1 | 63 | 24 |
| Threonine | 120.1 | 74.2 | 63 | 8 |
| Threonine | 120.1 | 56 | 63 | 0 |
| Tryptophane | 205.1 | 188.1 | 63 | 4 |
| Tryptophane | 205.1 | 146 | 63 | 8 |
| Tryptophane | 205.1 | 118.1 | 63 | 28 |
| Tyrosine | 182.1 | 136.2 | 63 | 12 |
| Tyrosine | 182.1 | 123.1 | 63 | 0 |
| Valine | 118.1 | 72.2 | 63 | 8 |
| Valine | 118.1 | 55.1 | 63 | 0 |

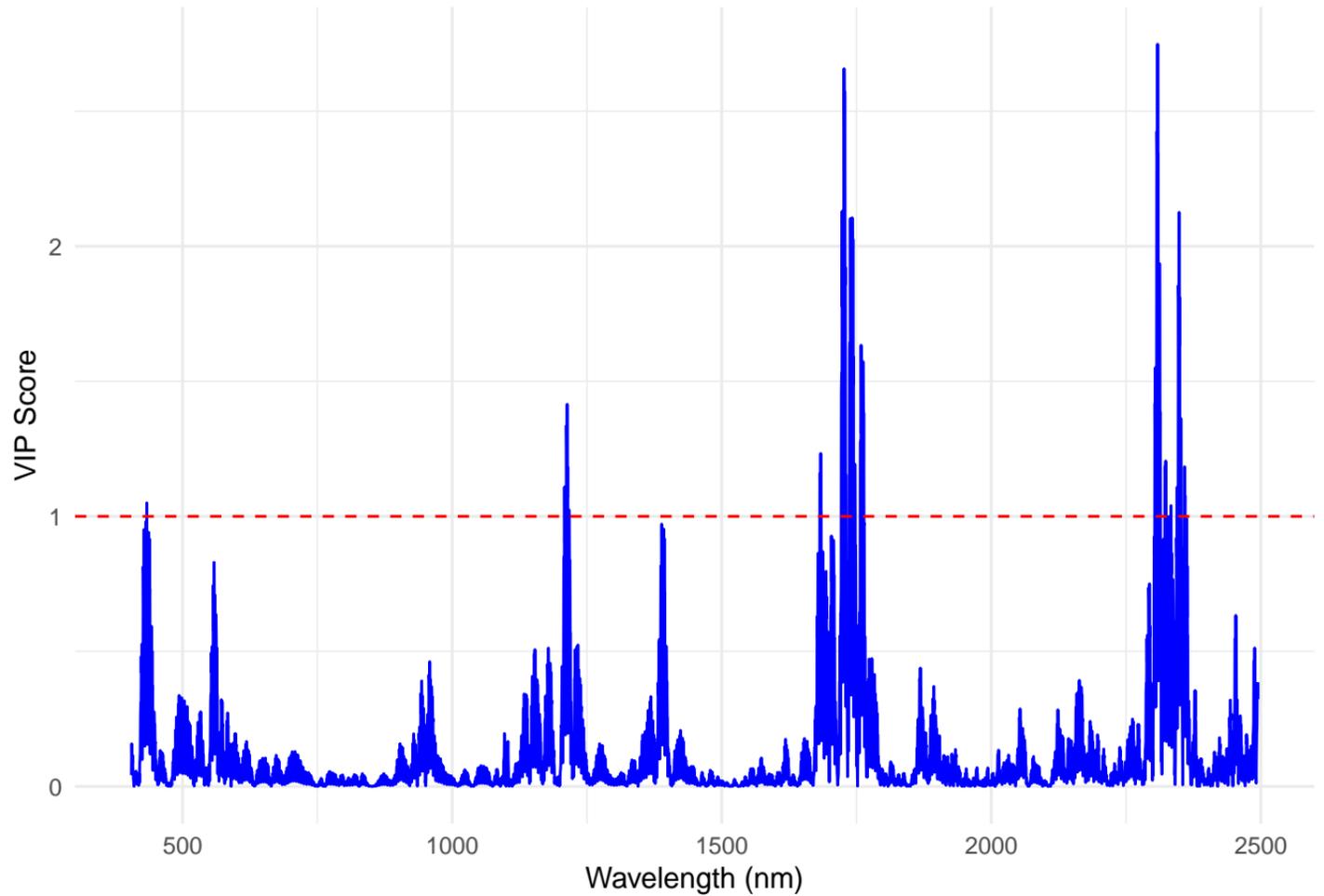

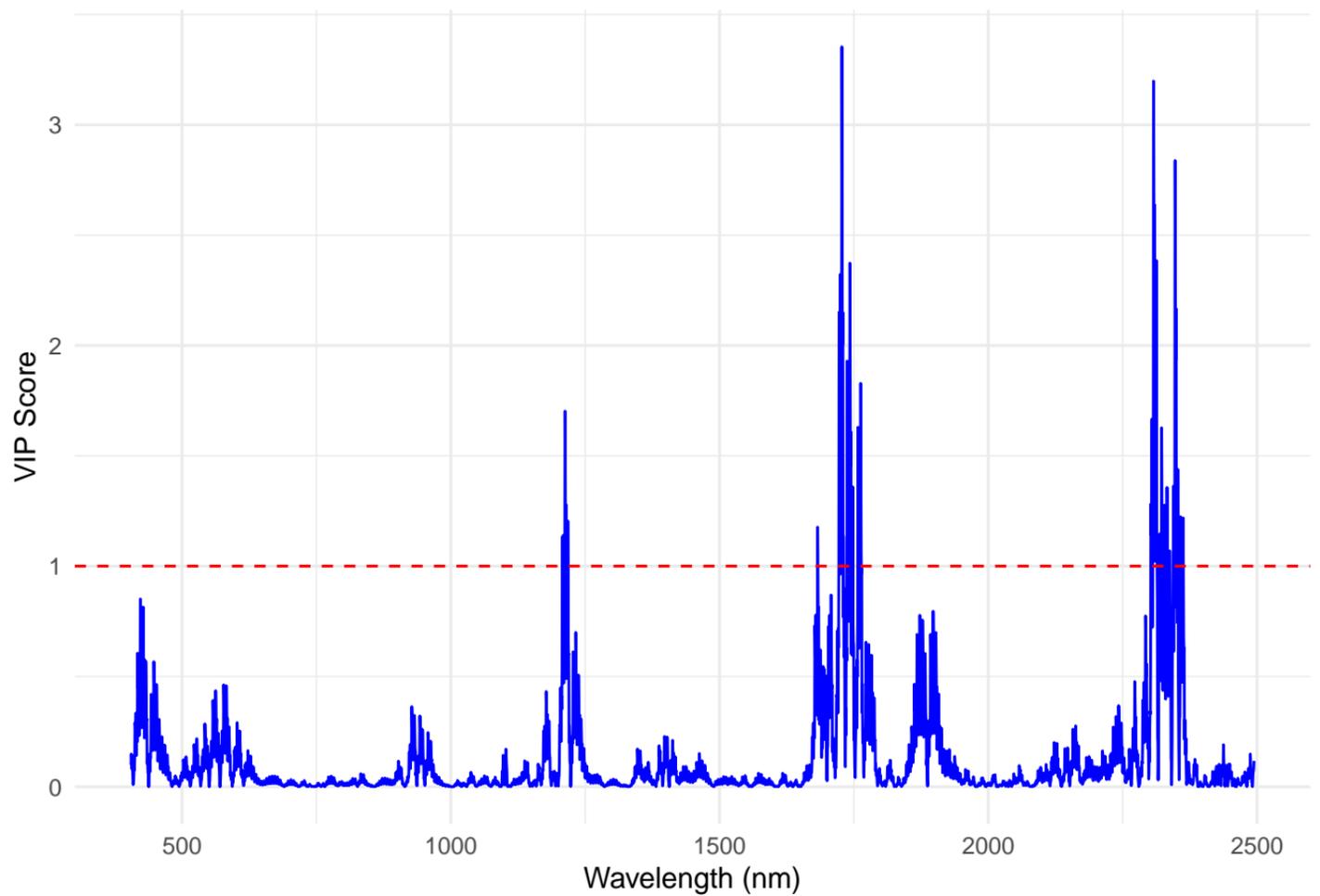

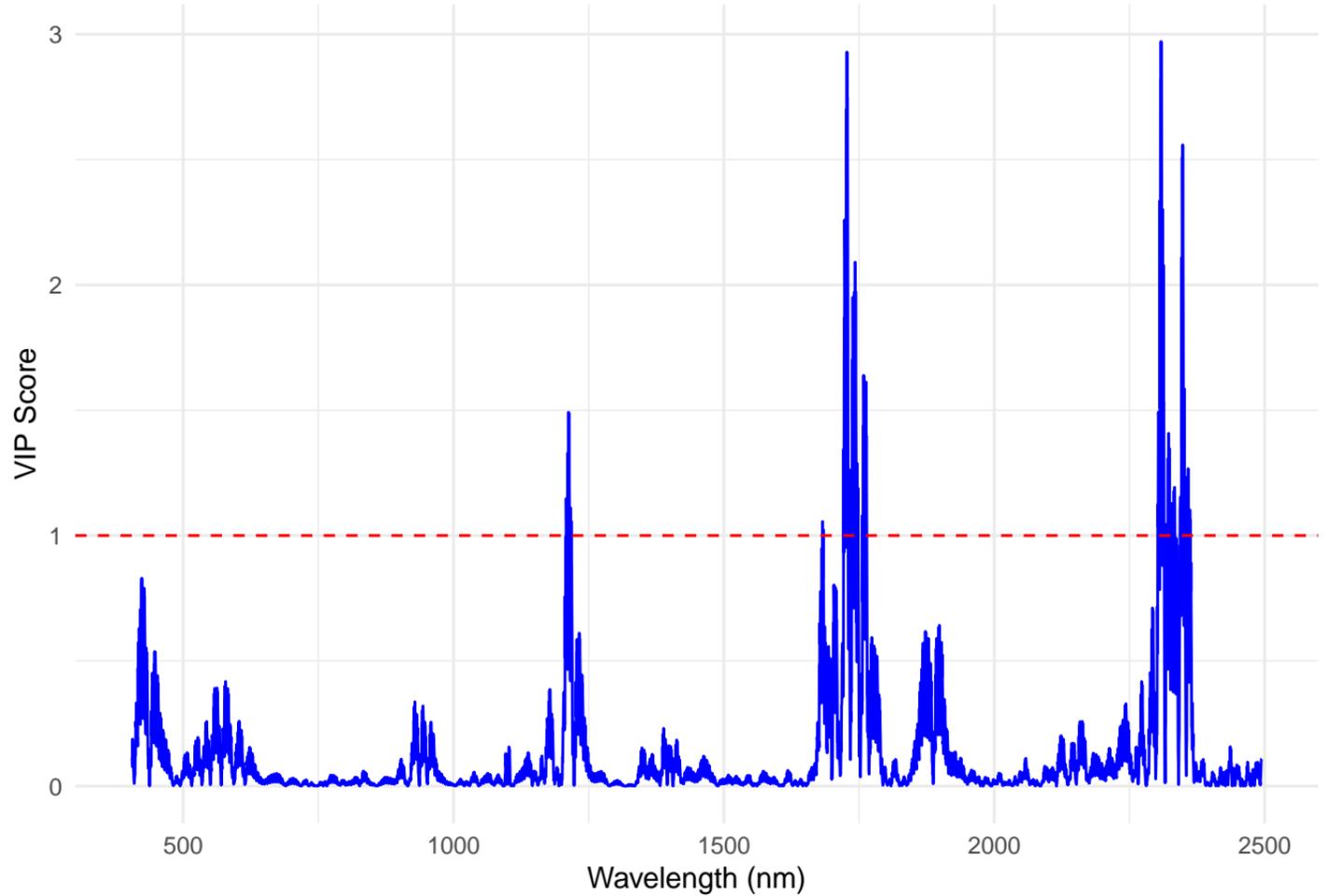

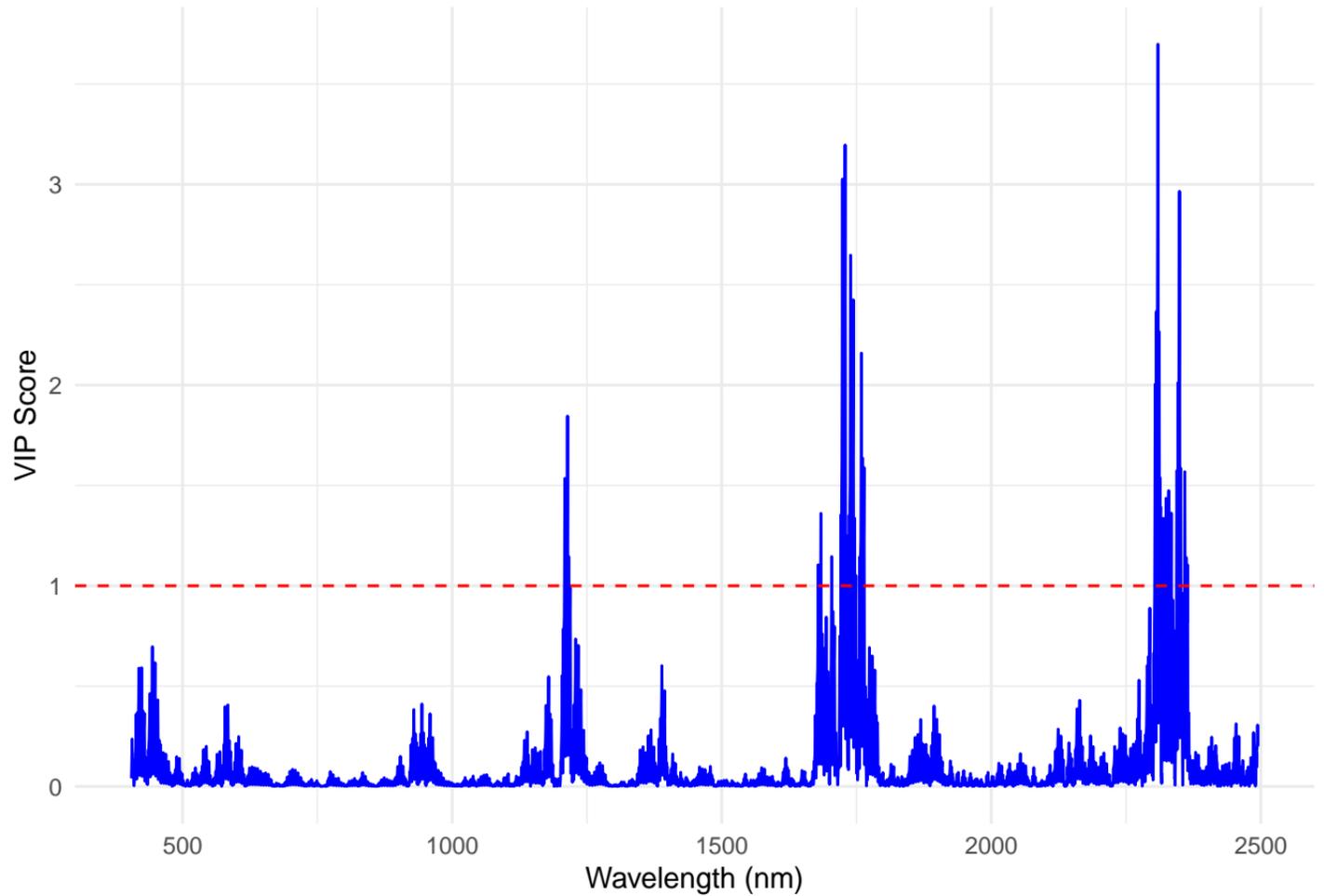

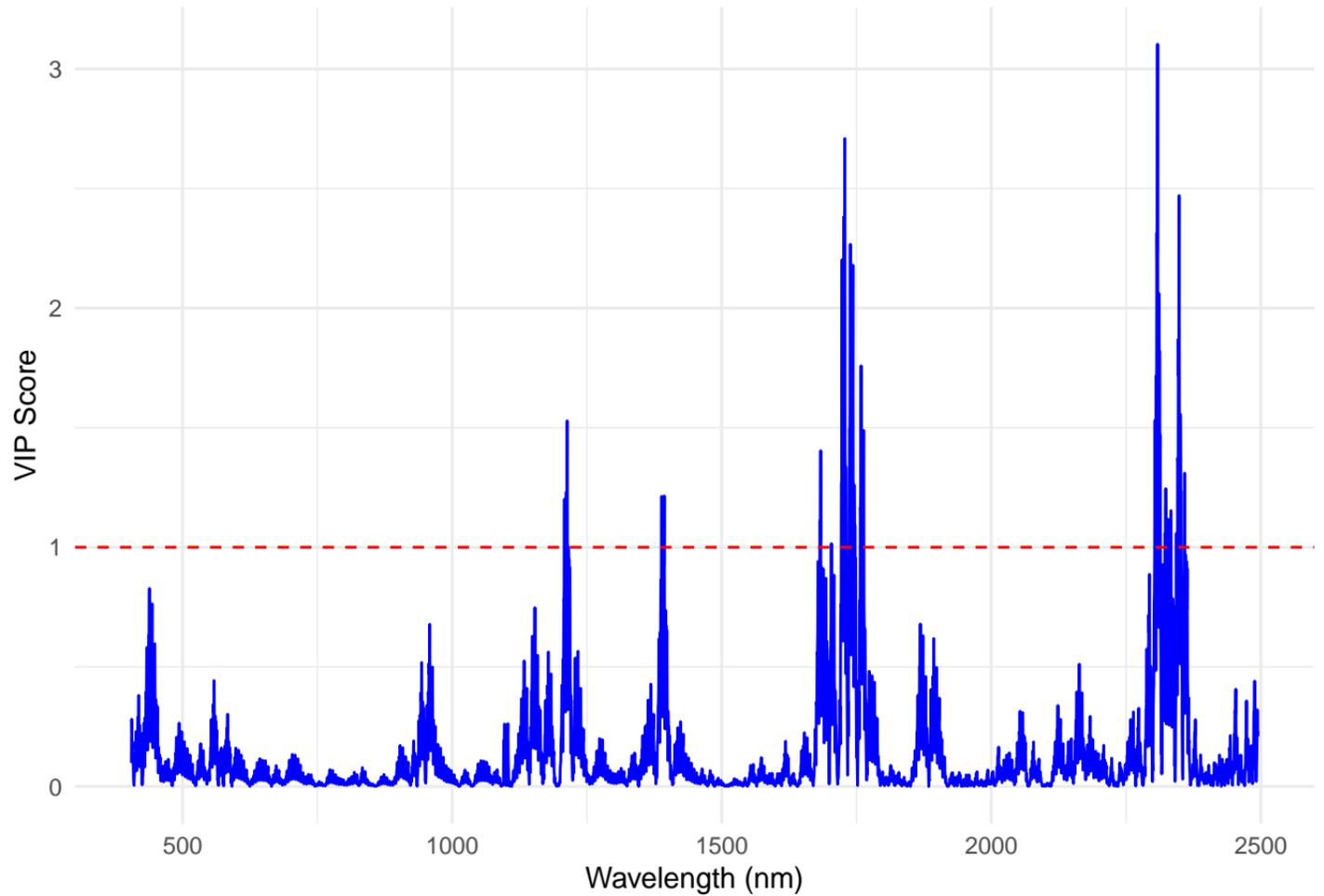

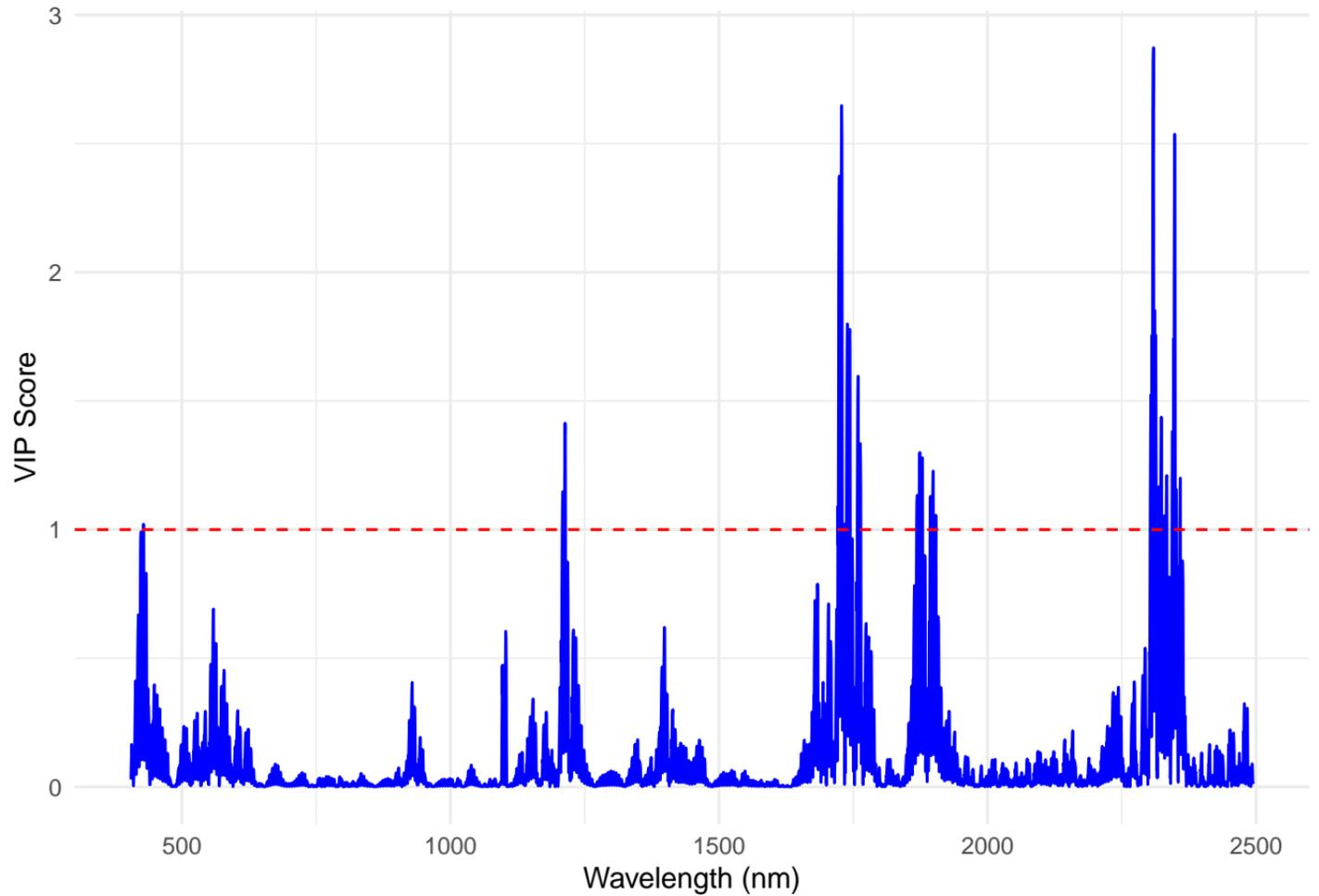

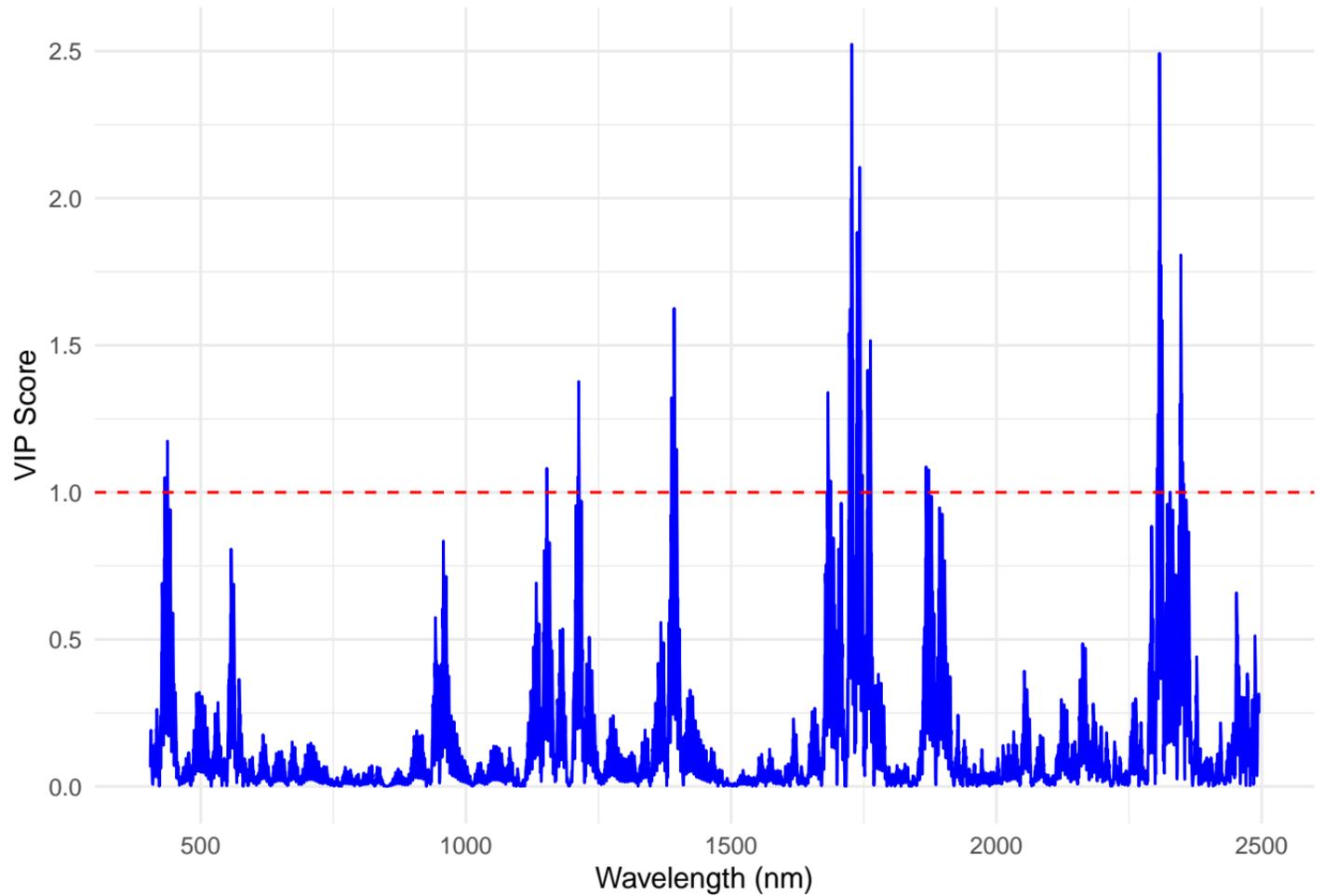

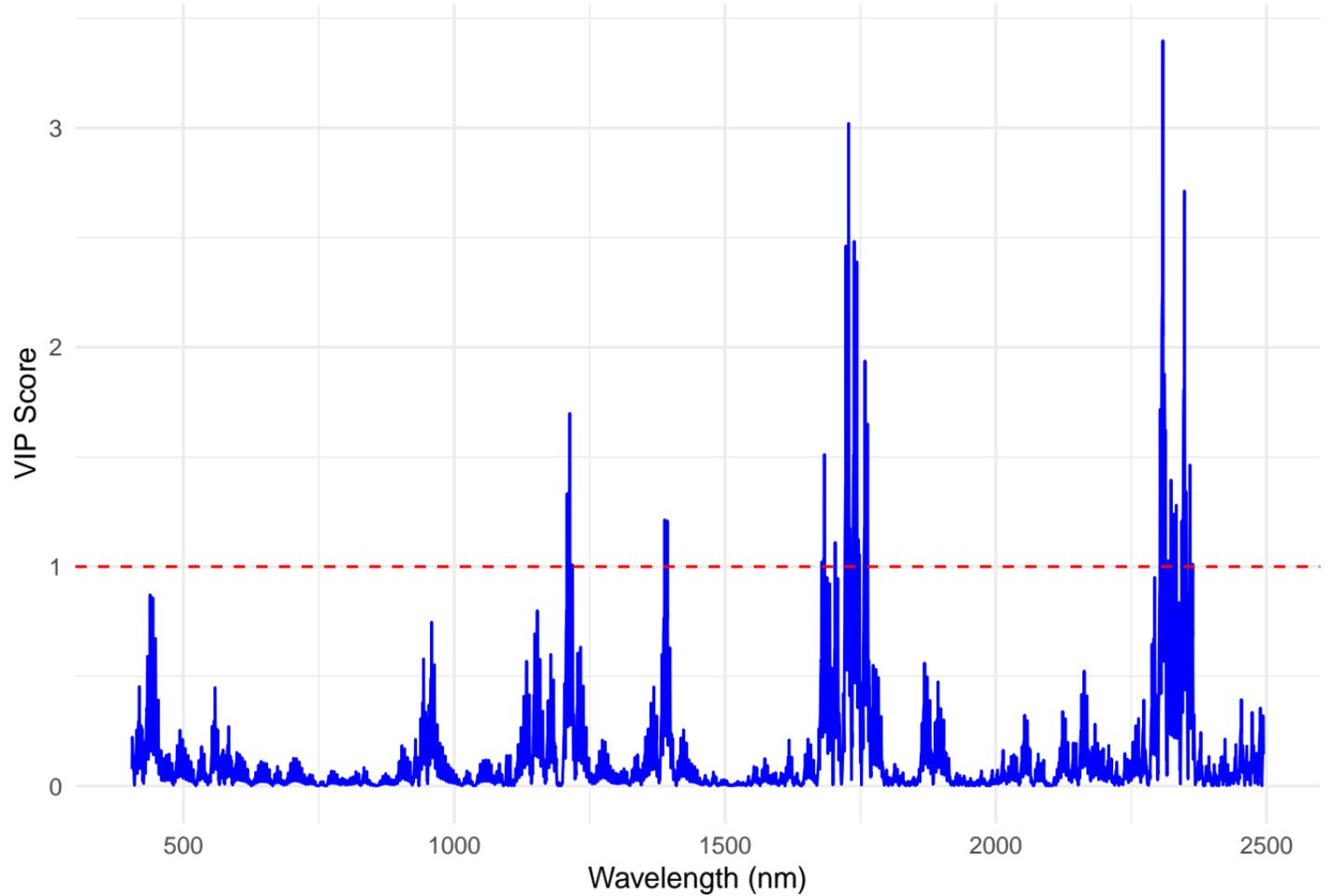

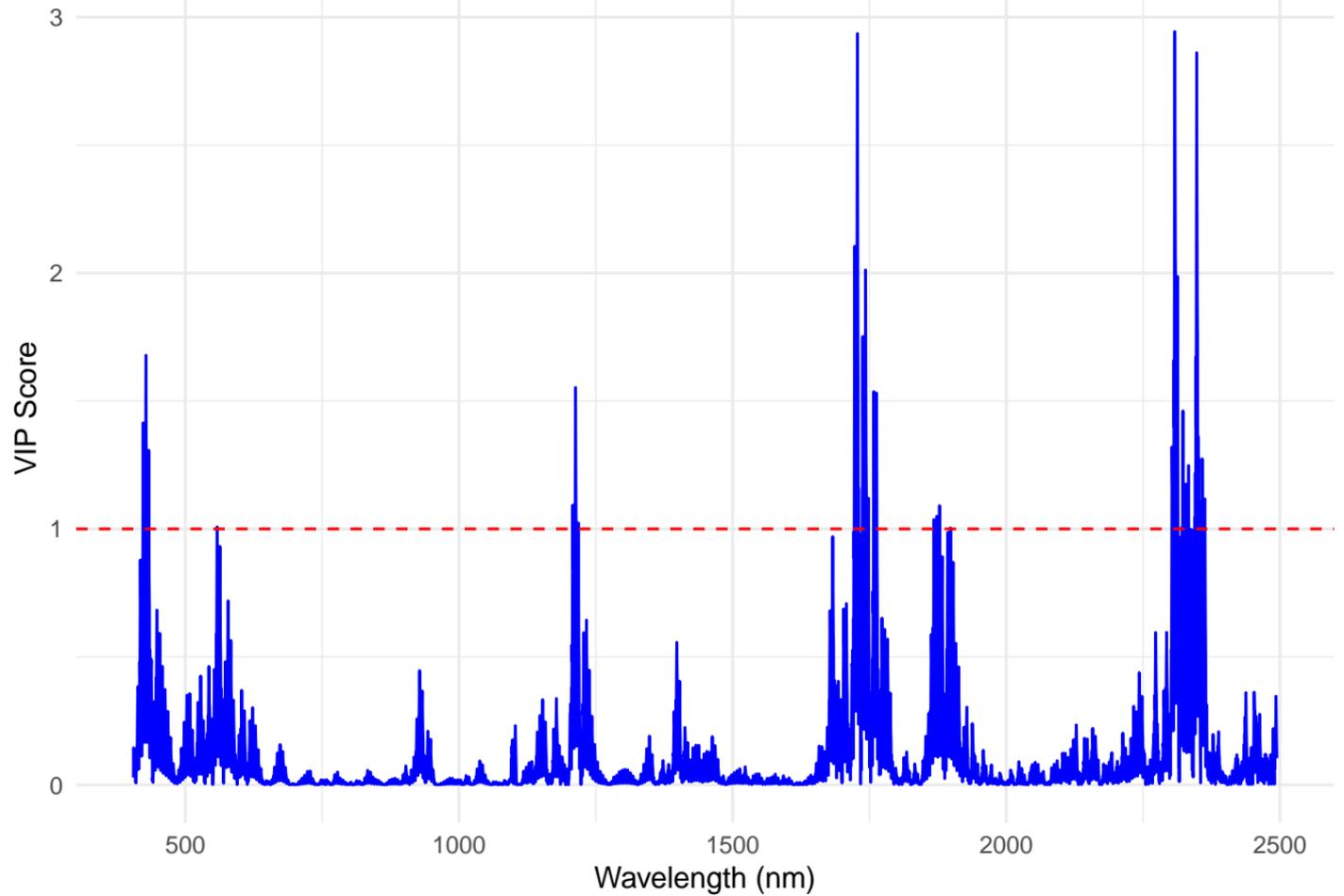

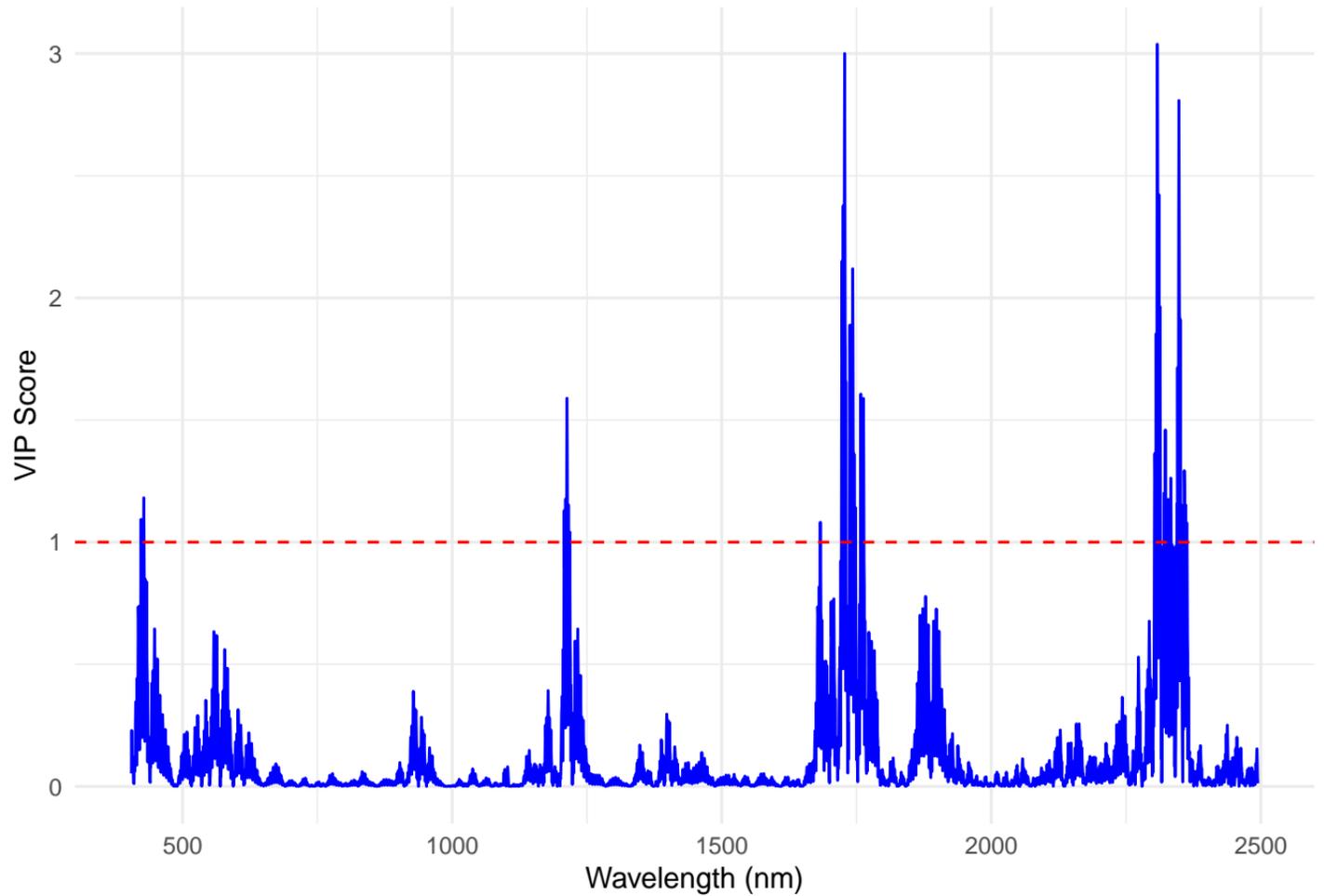

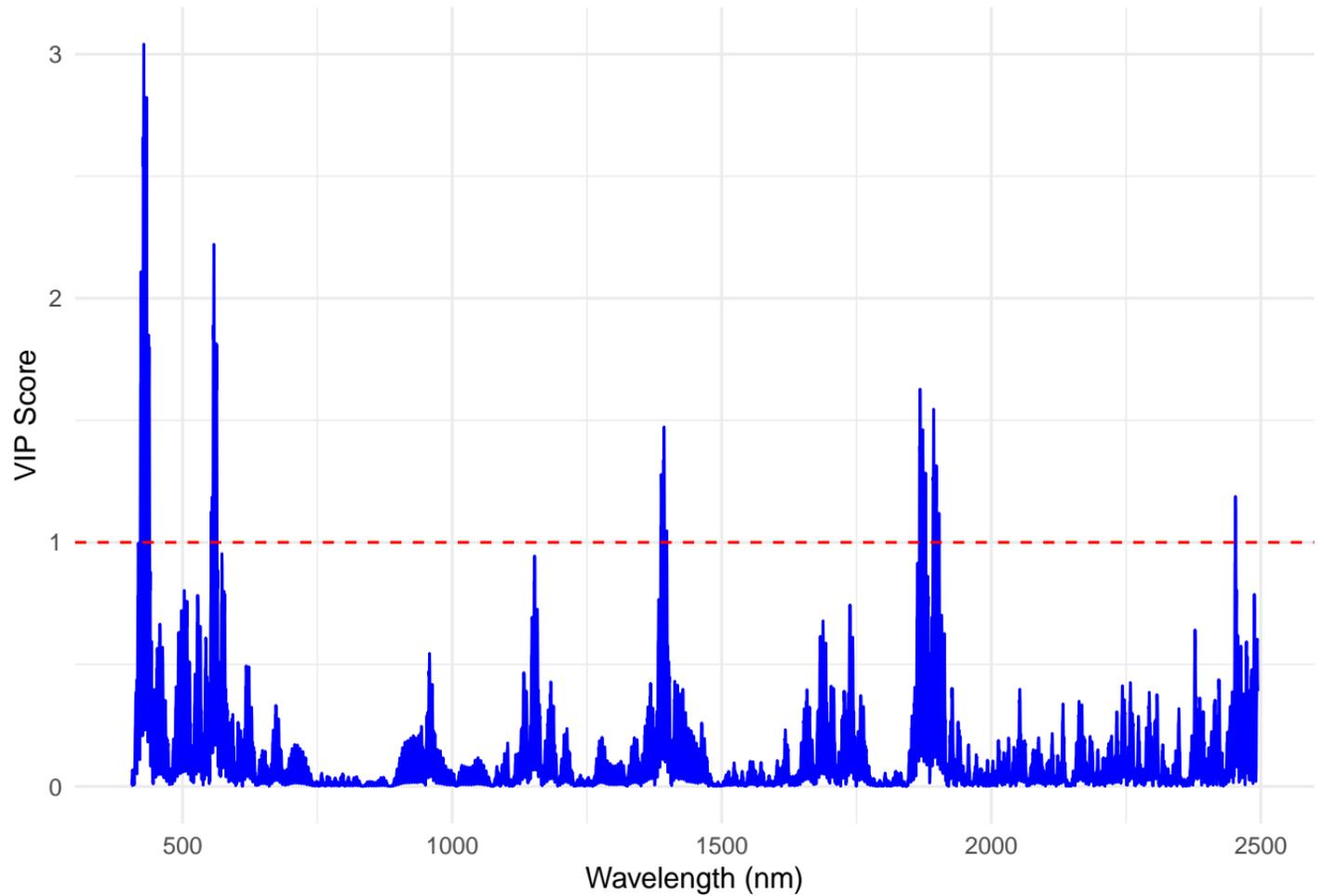

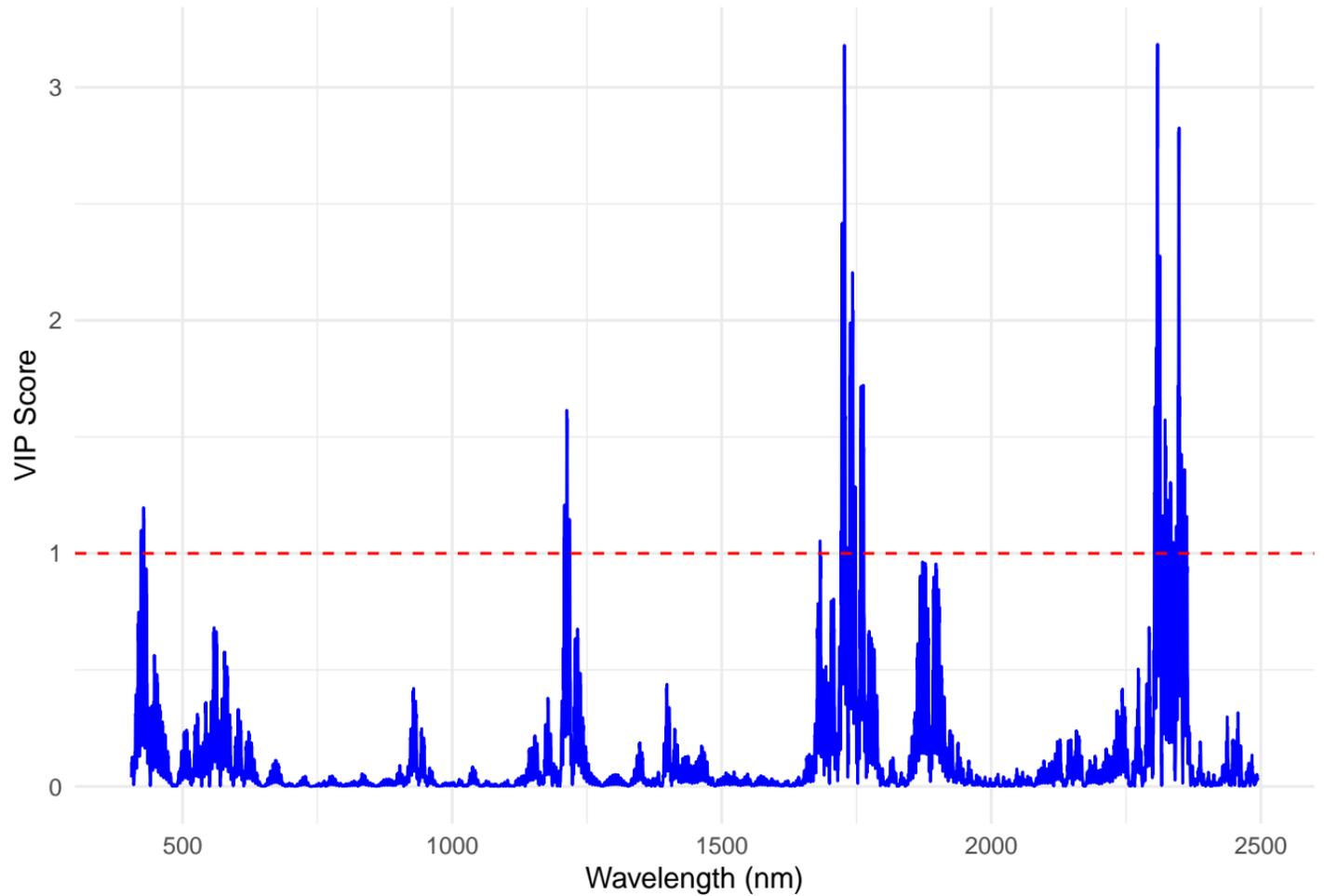

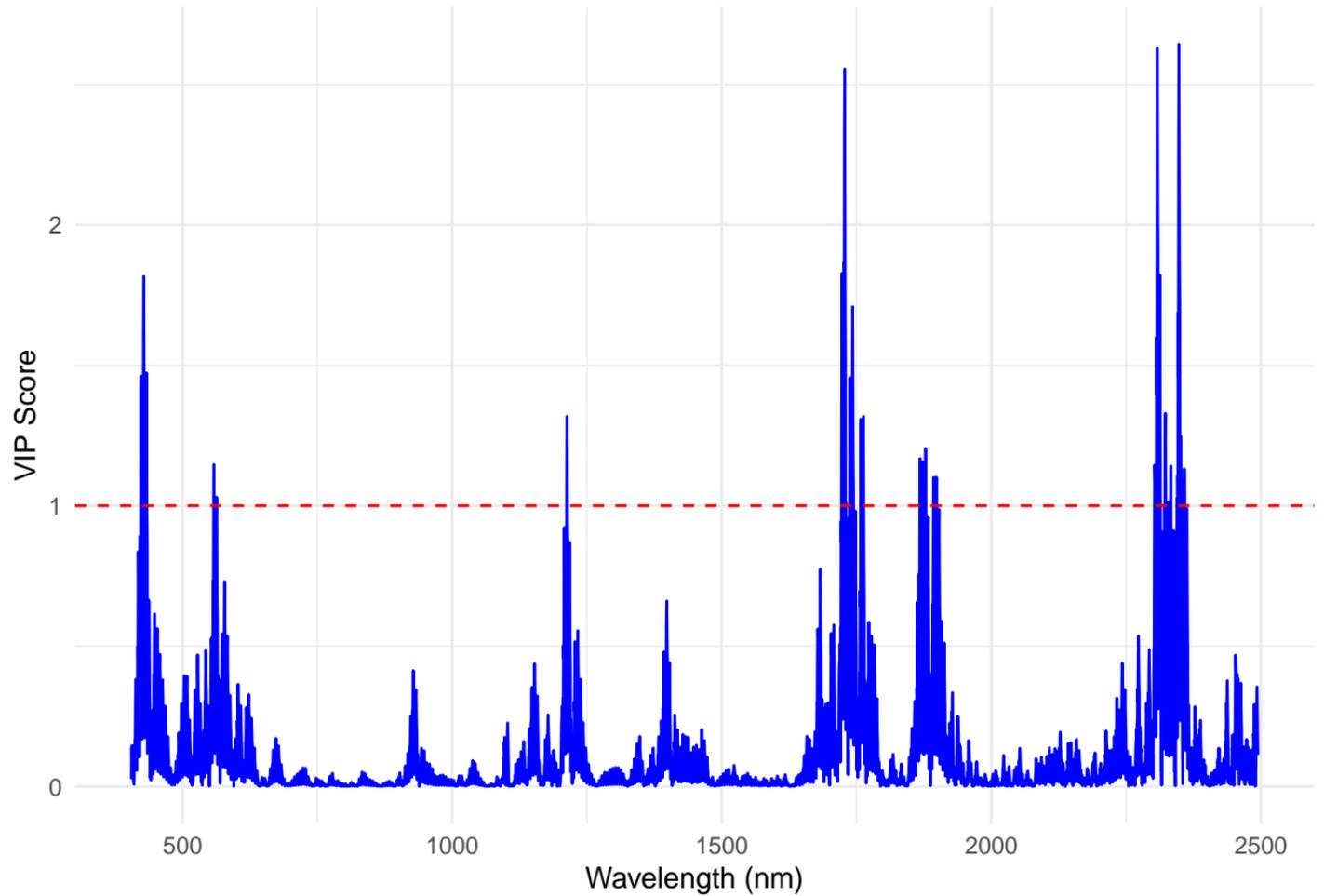

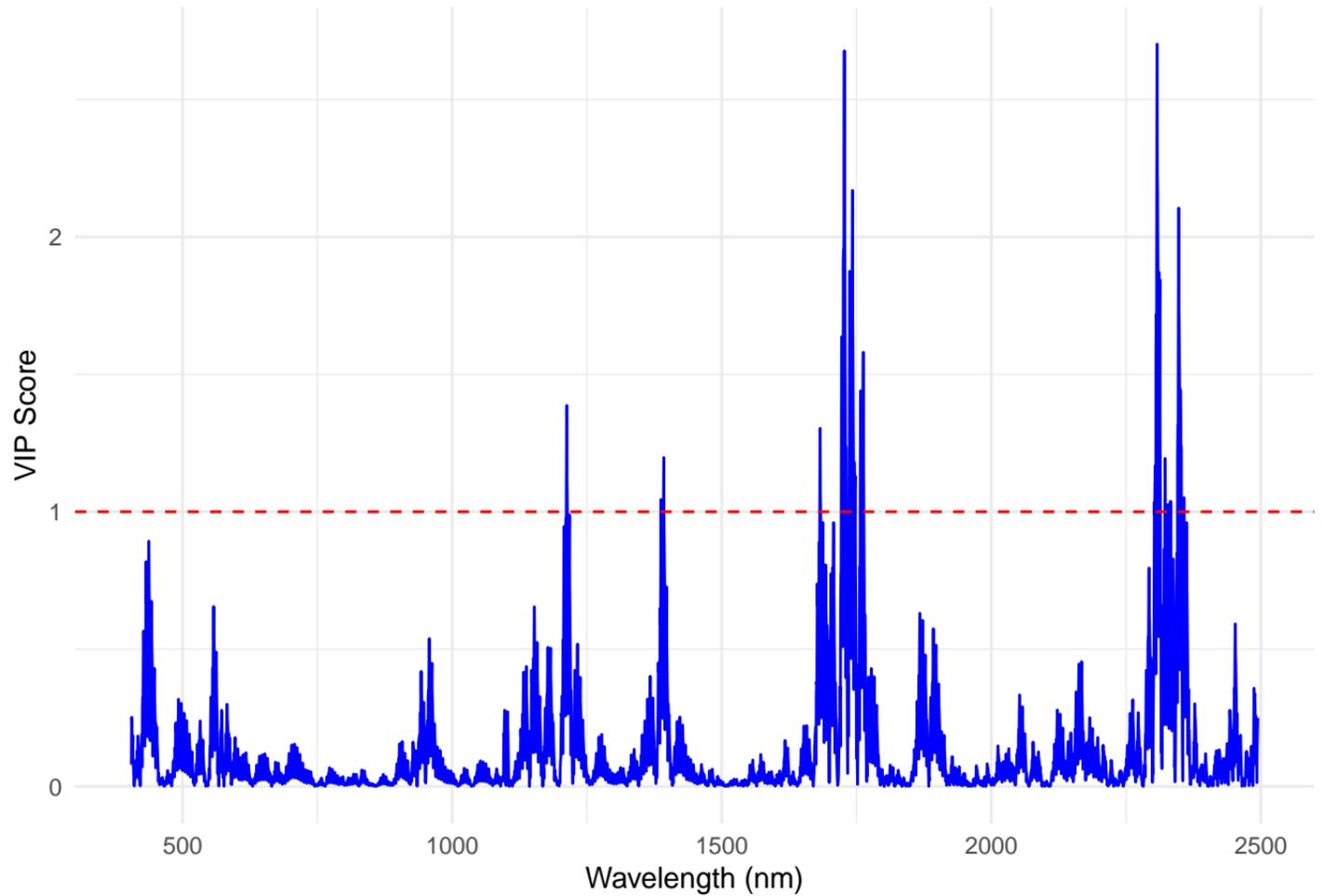

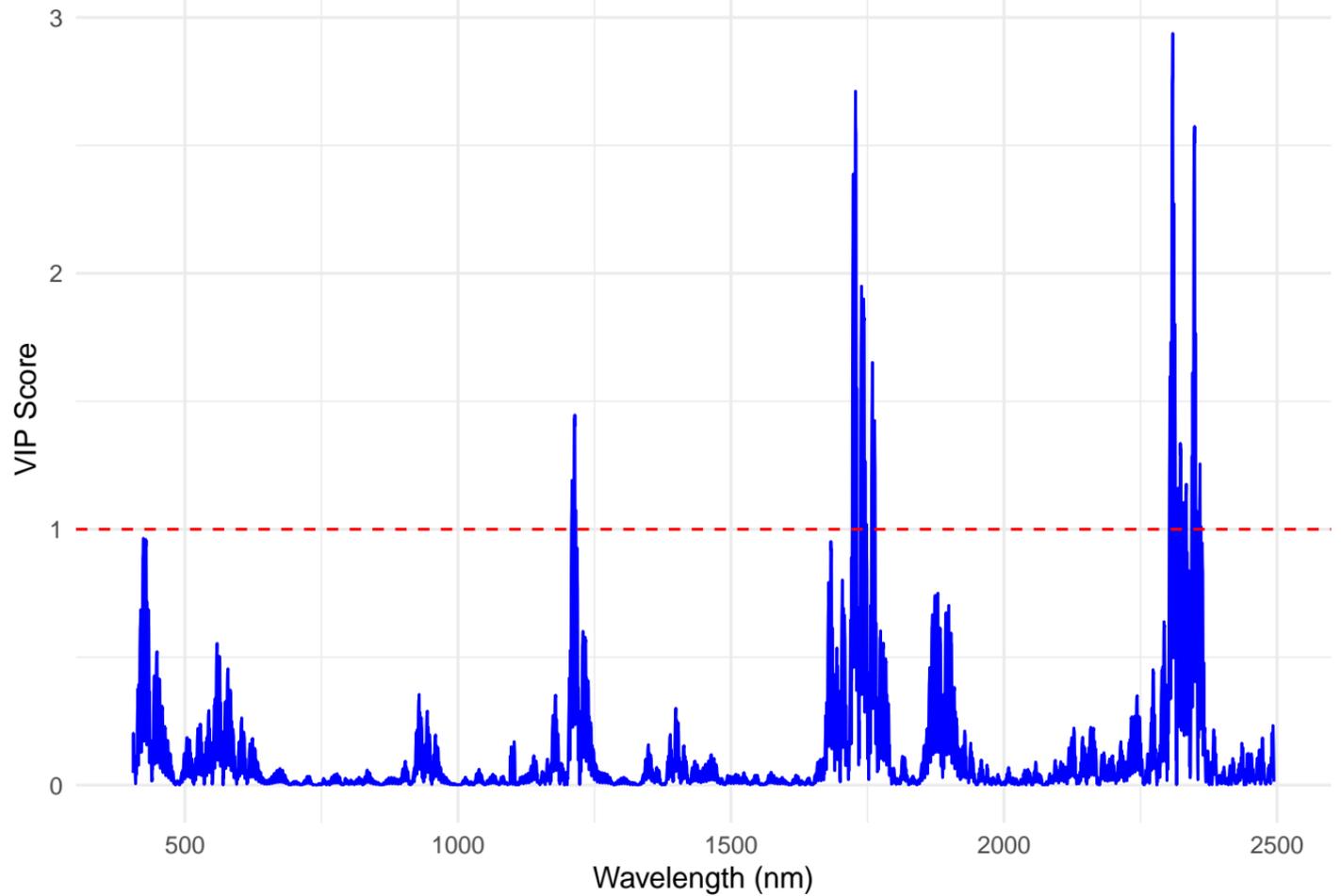

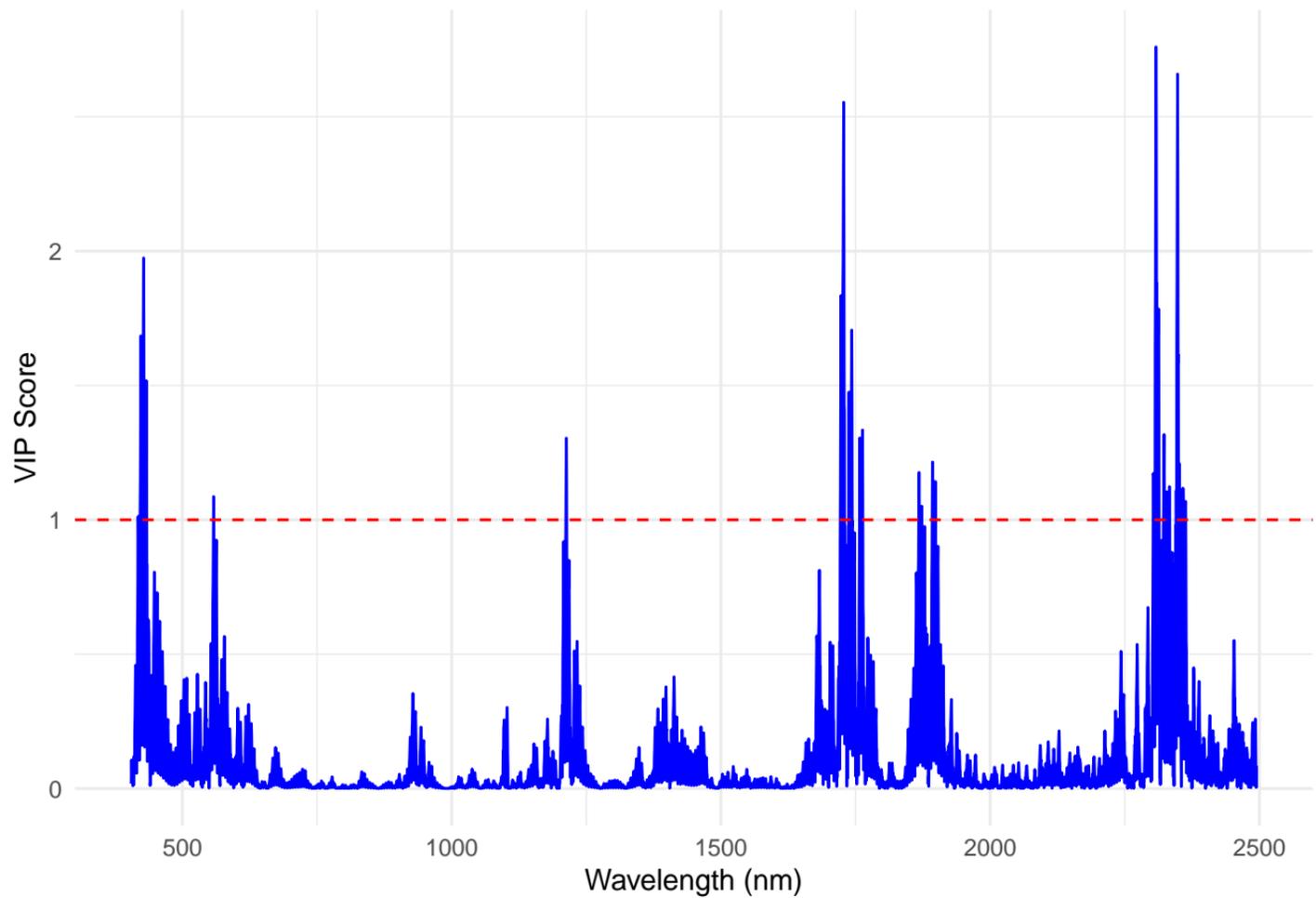

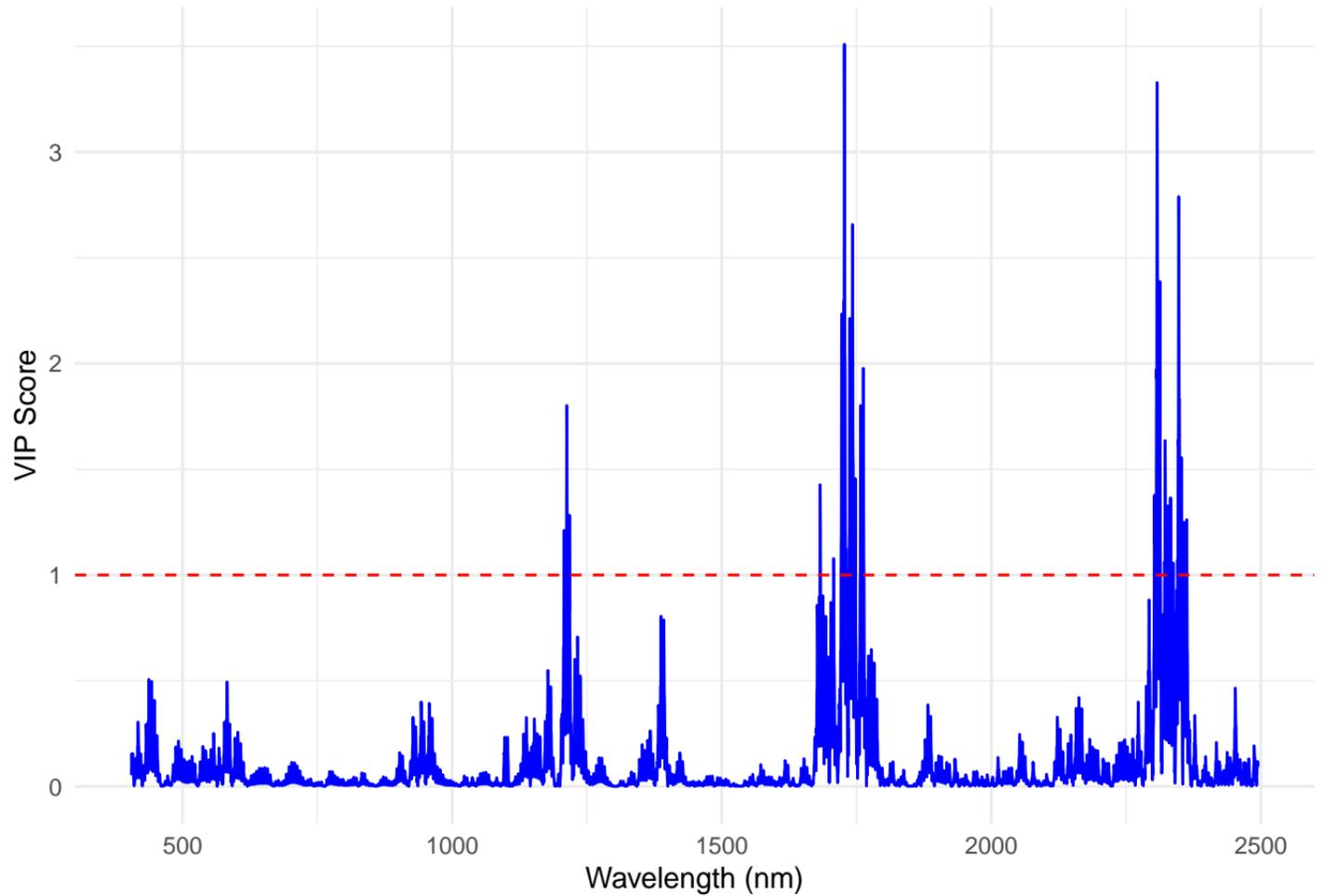

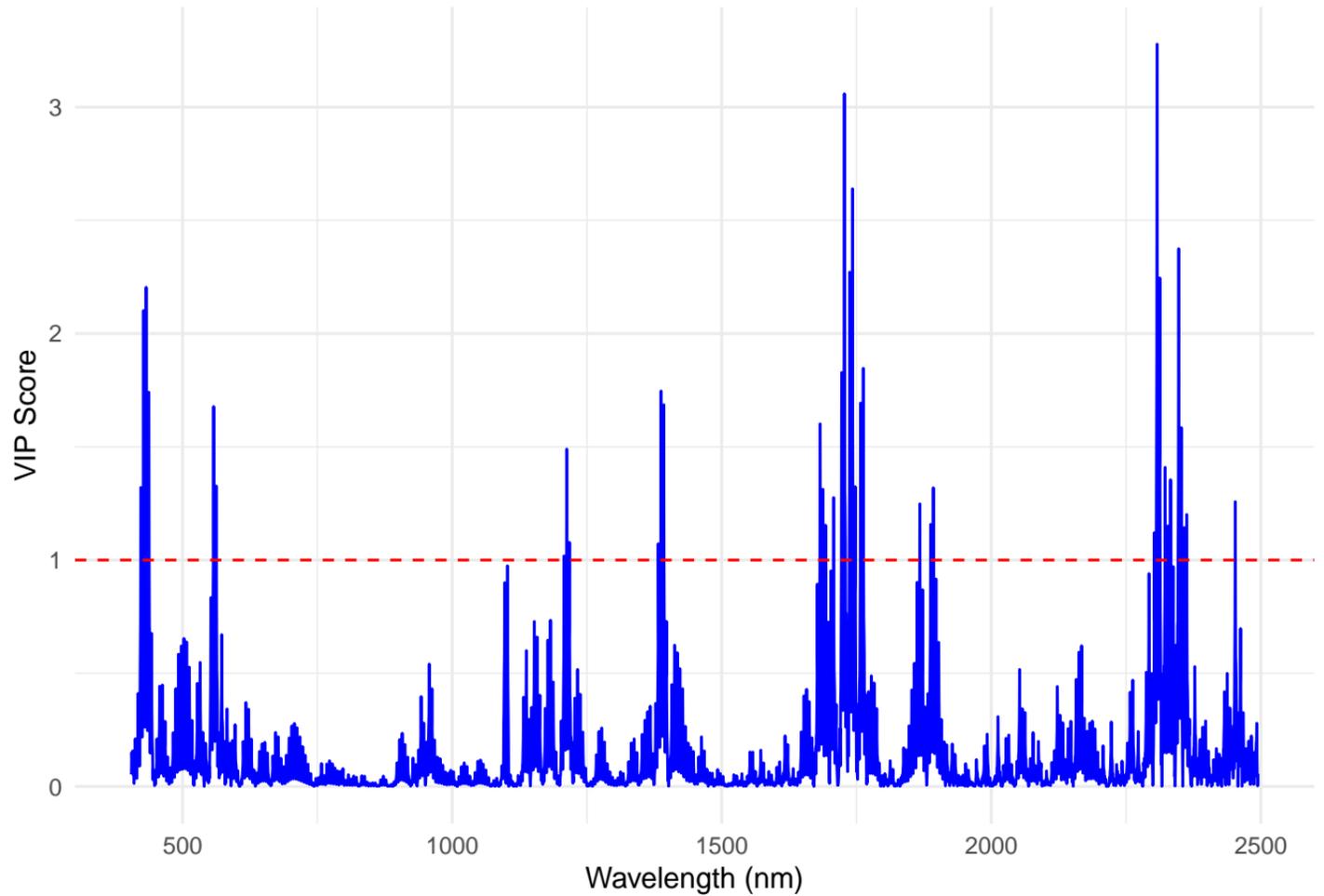

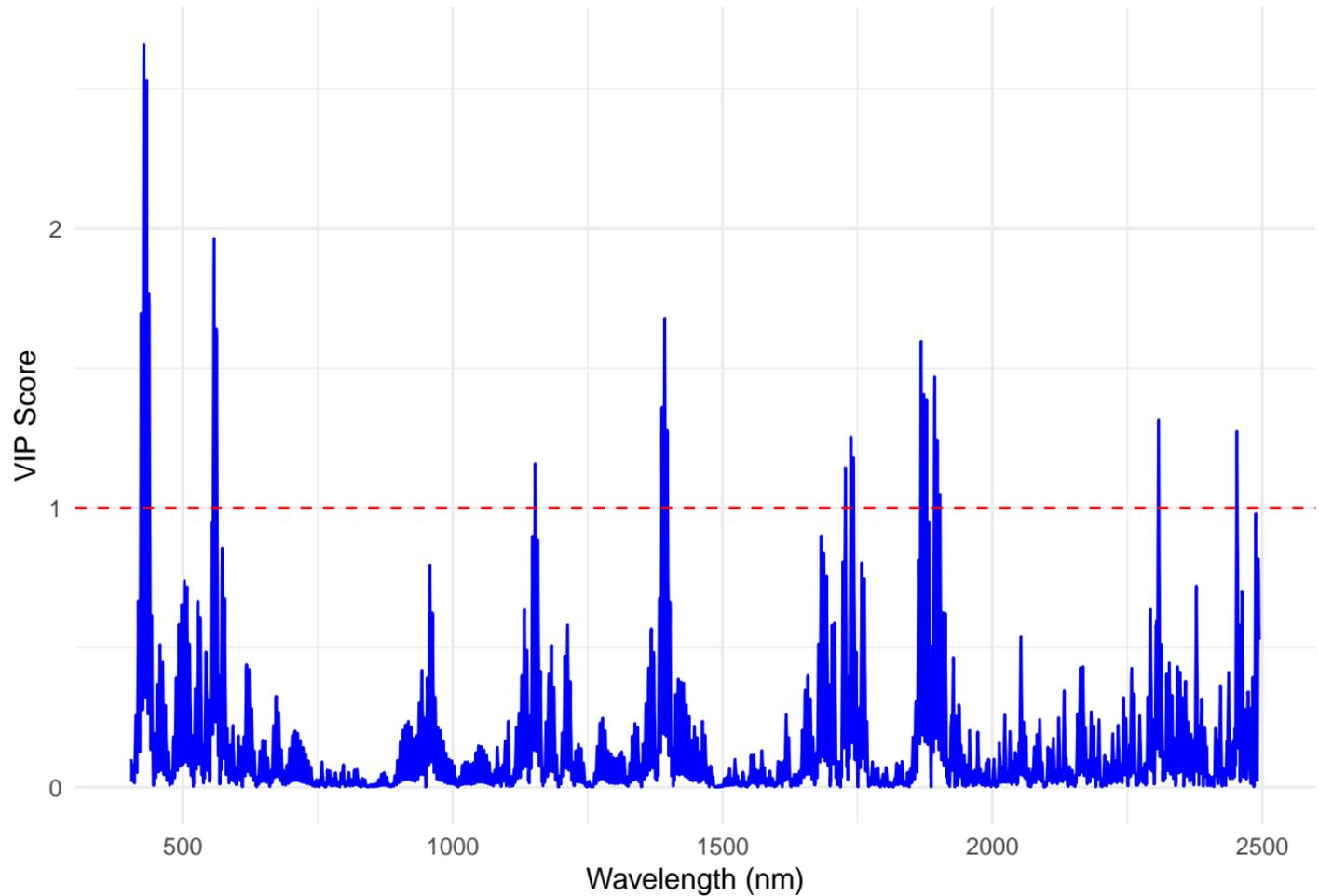

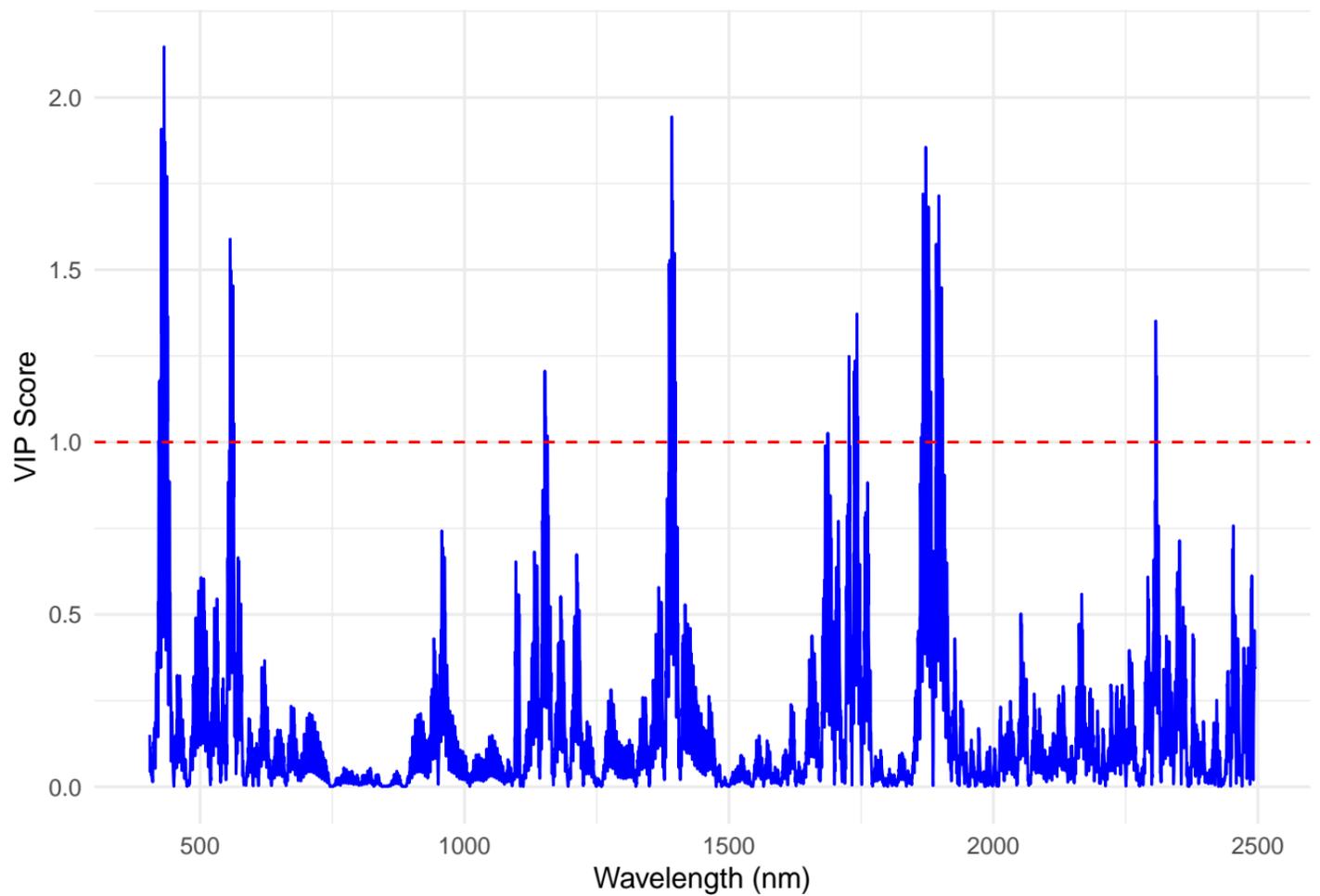

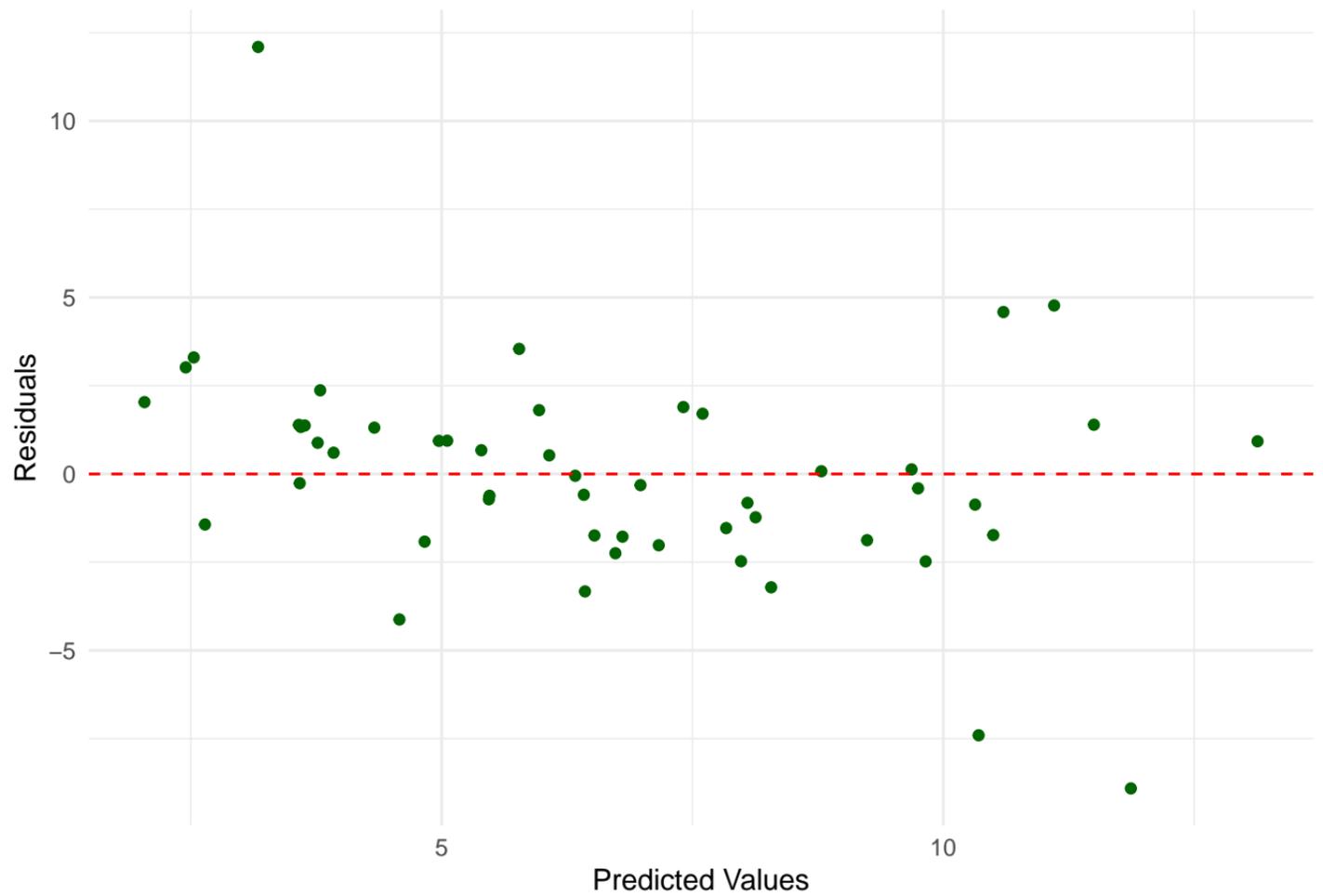

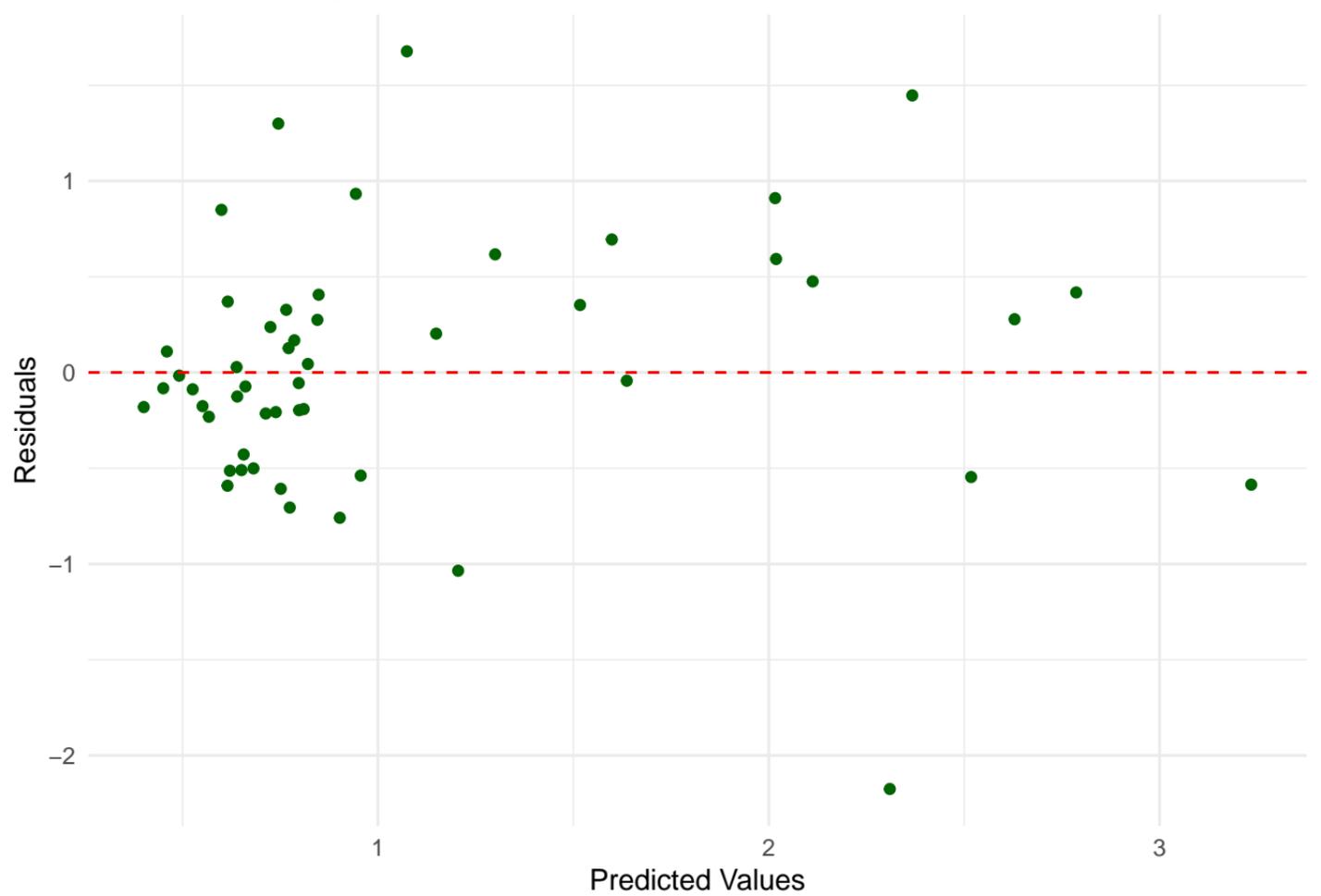

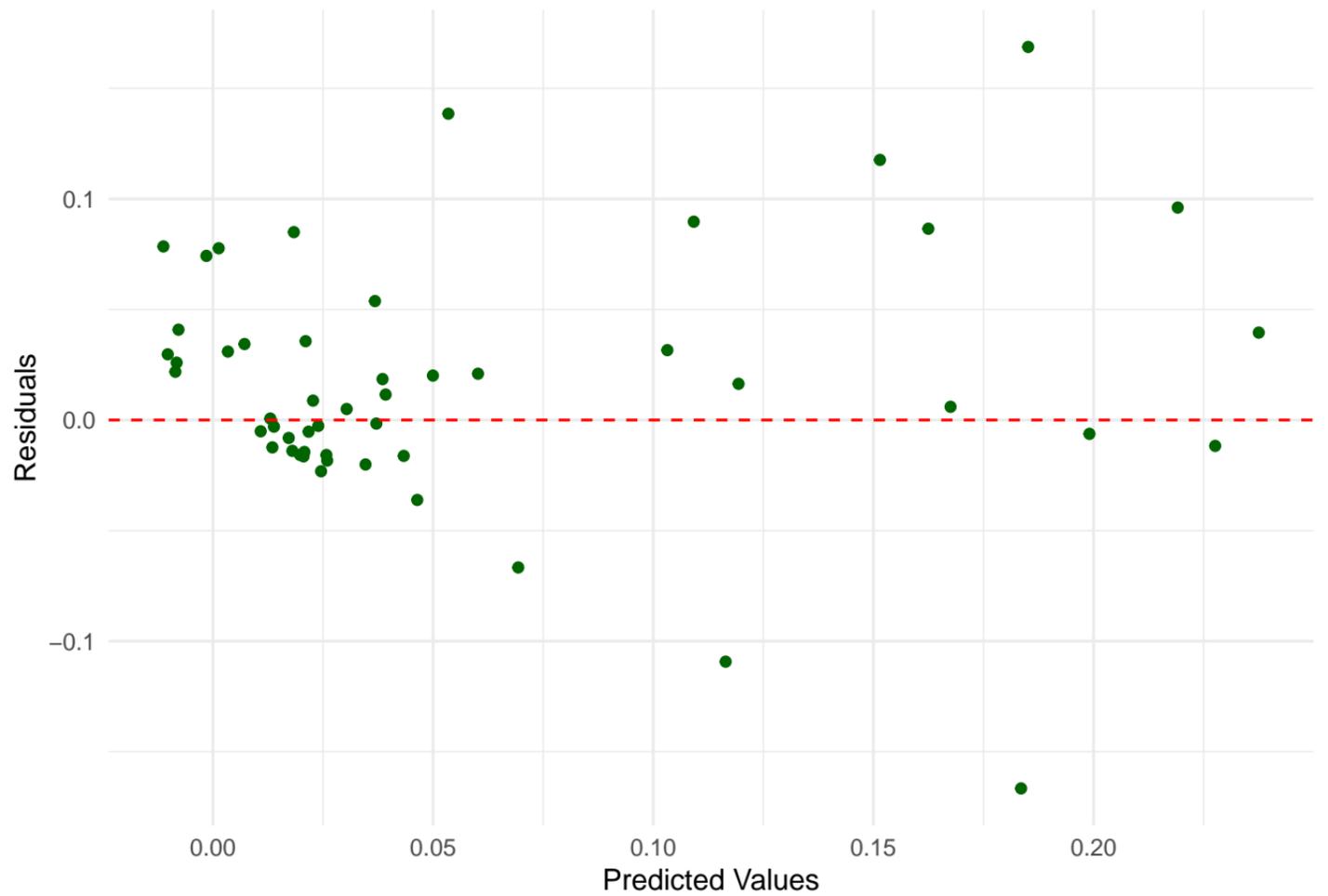

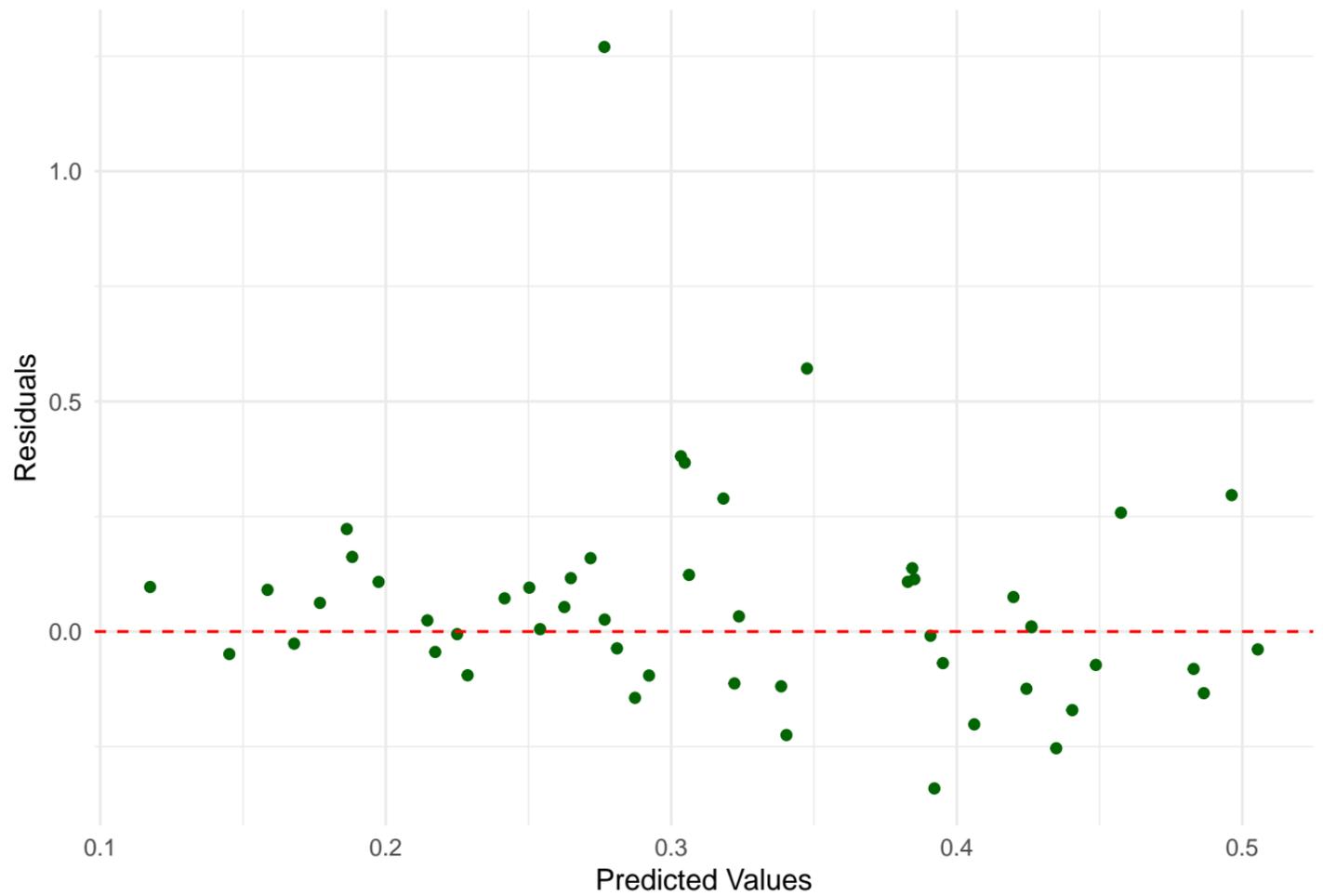

## Residuals for Glutamate

![Residuals for Glutamate scatter plot showing residuals versus predicted values, with a red dashed horizontal line at y=0. Points are distributed across predicted values from 0 to ~5.]

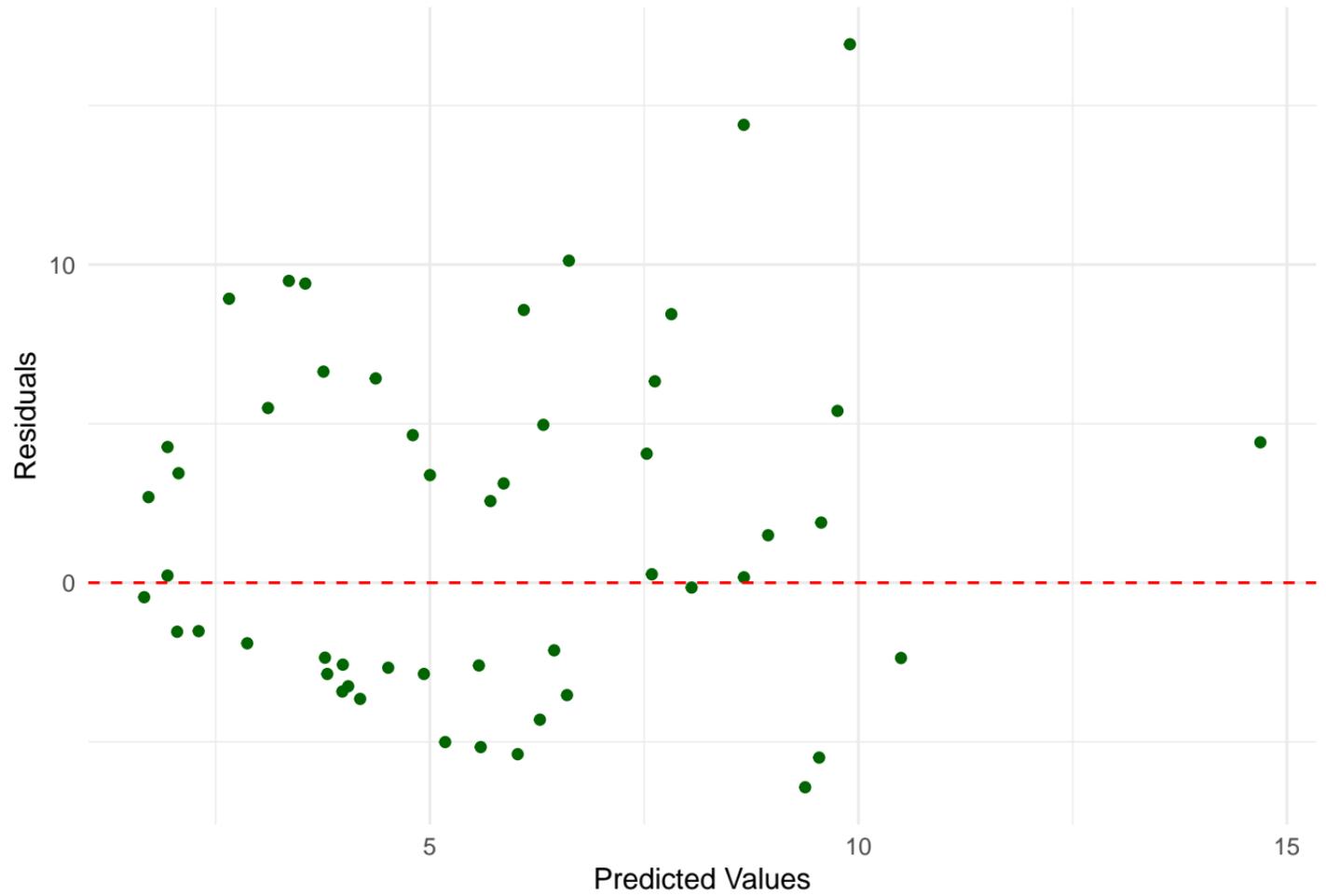

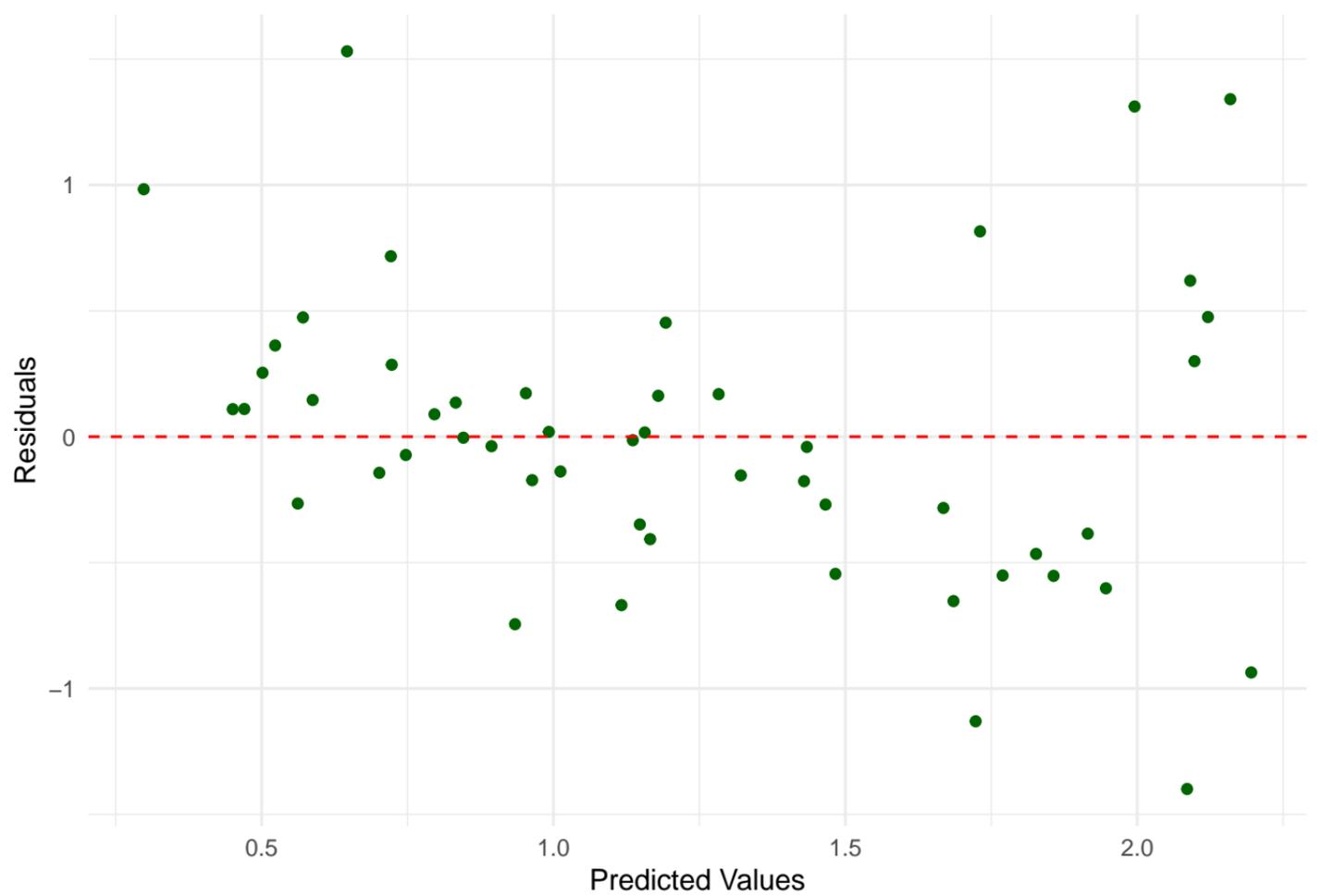

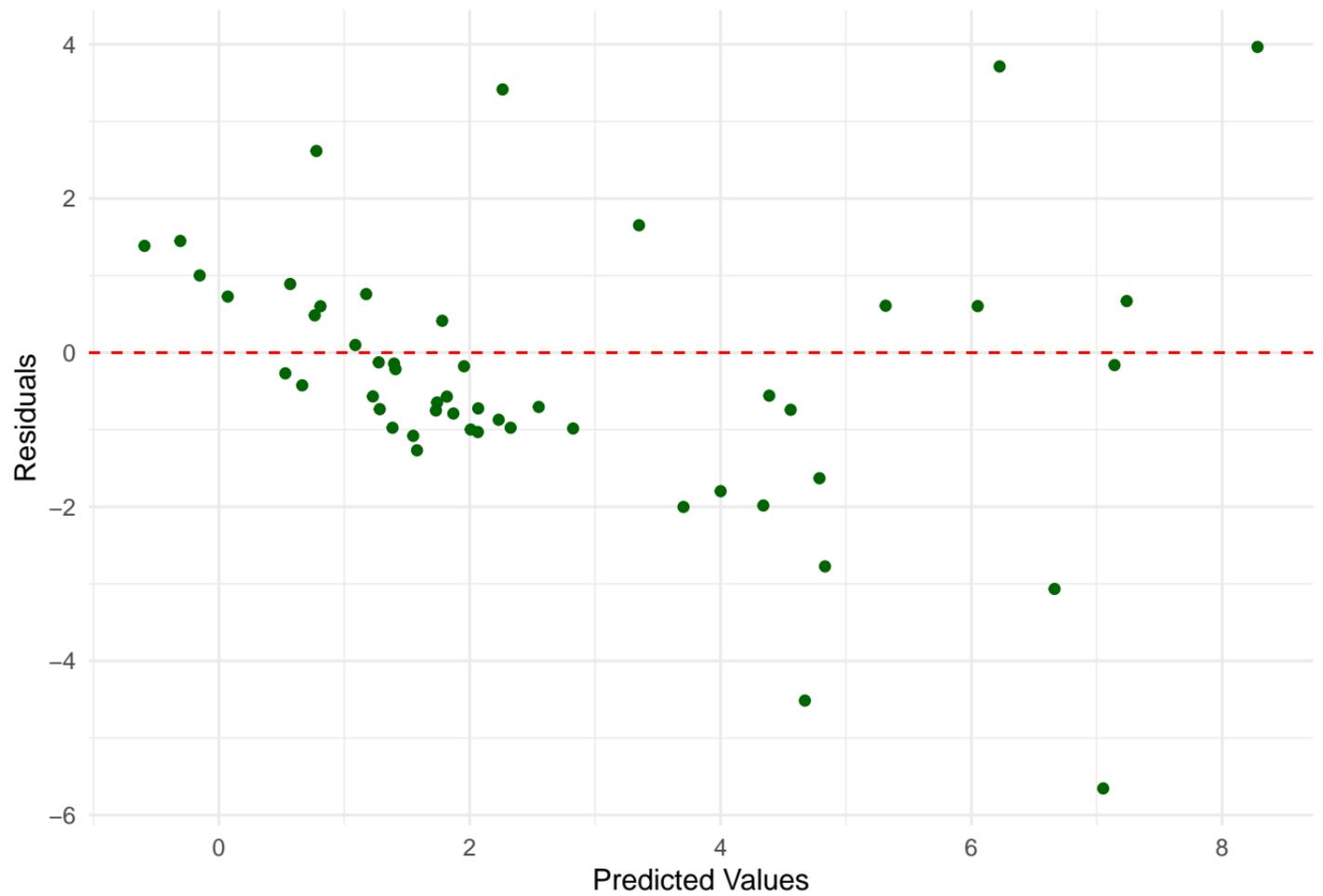

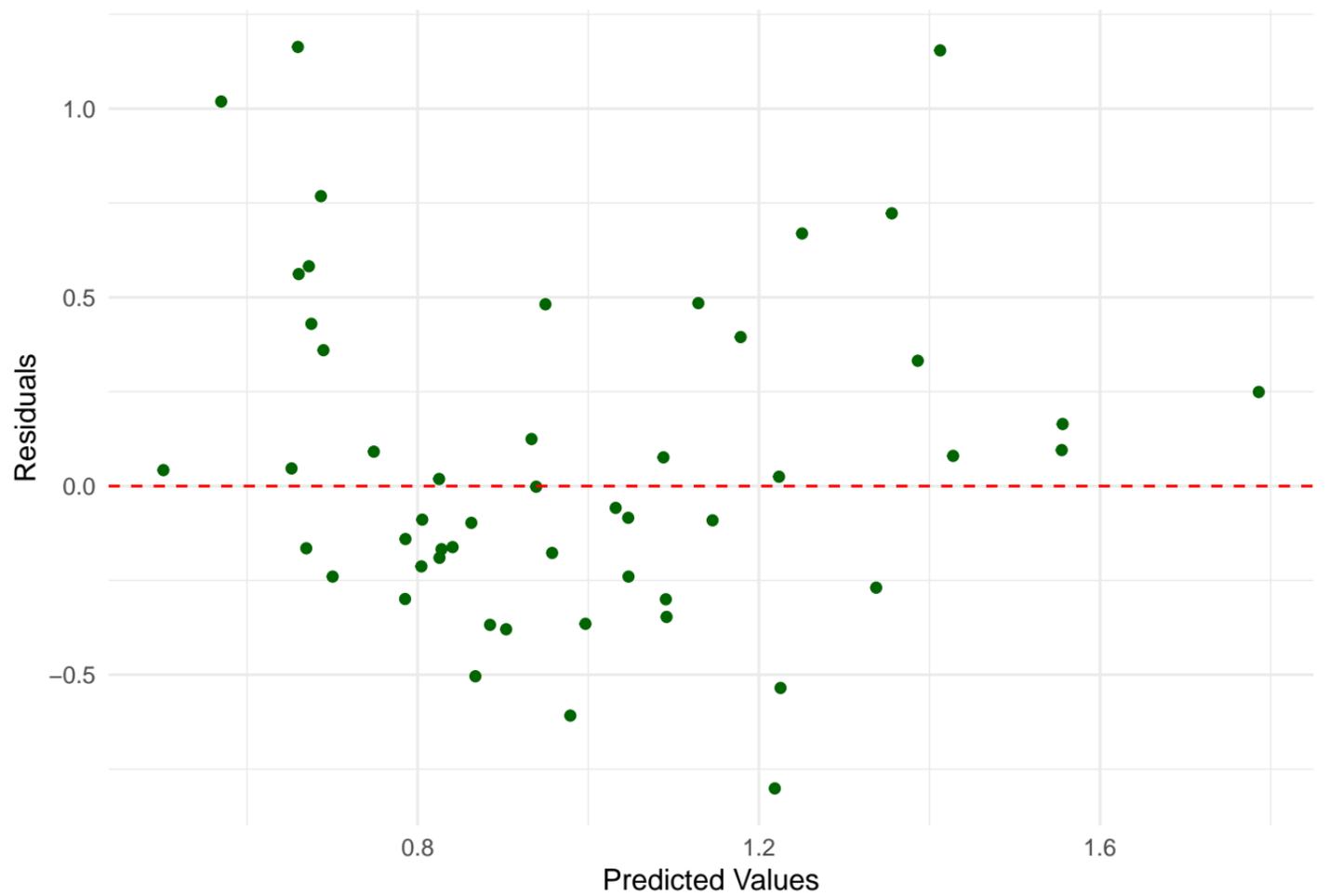

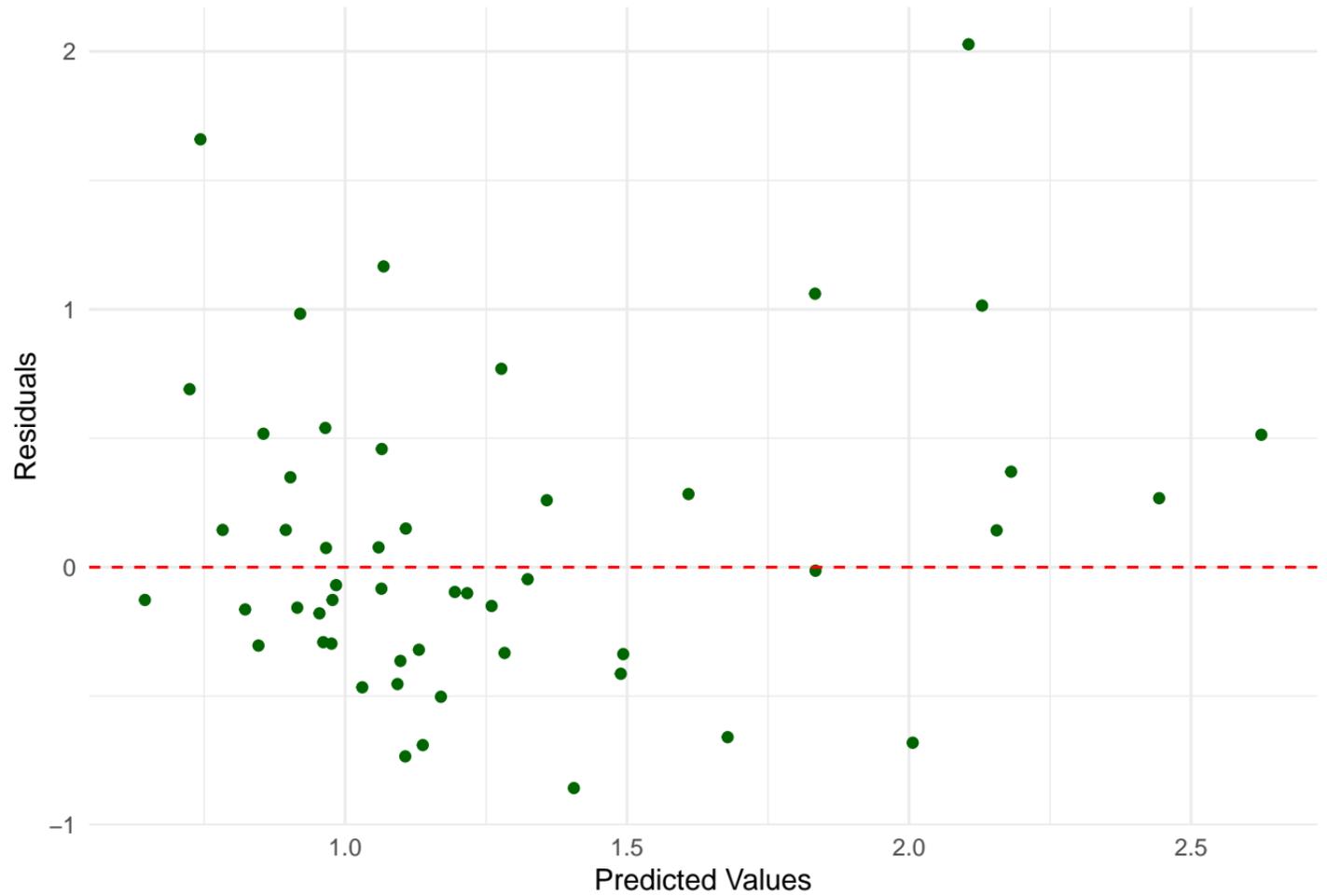

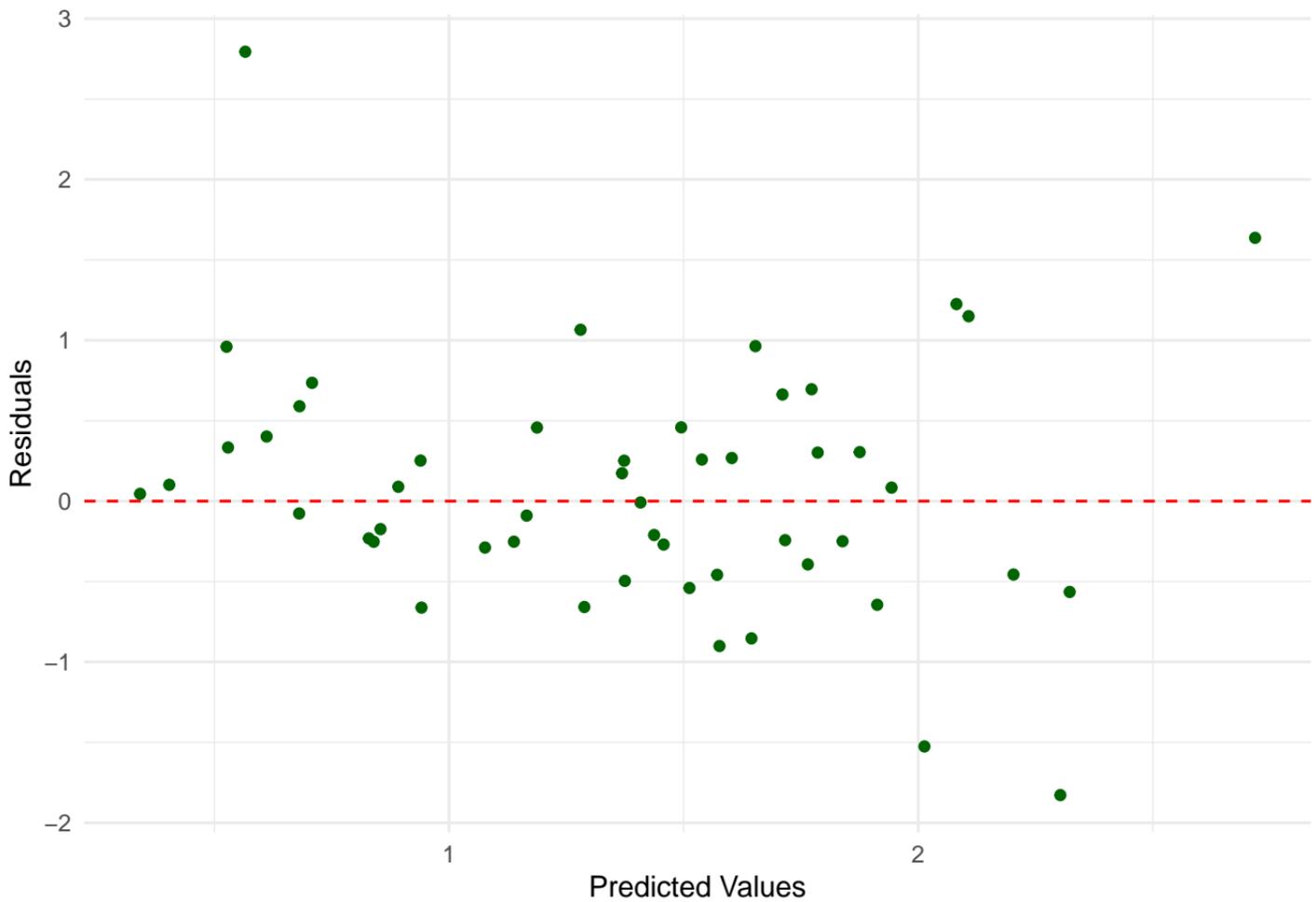

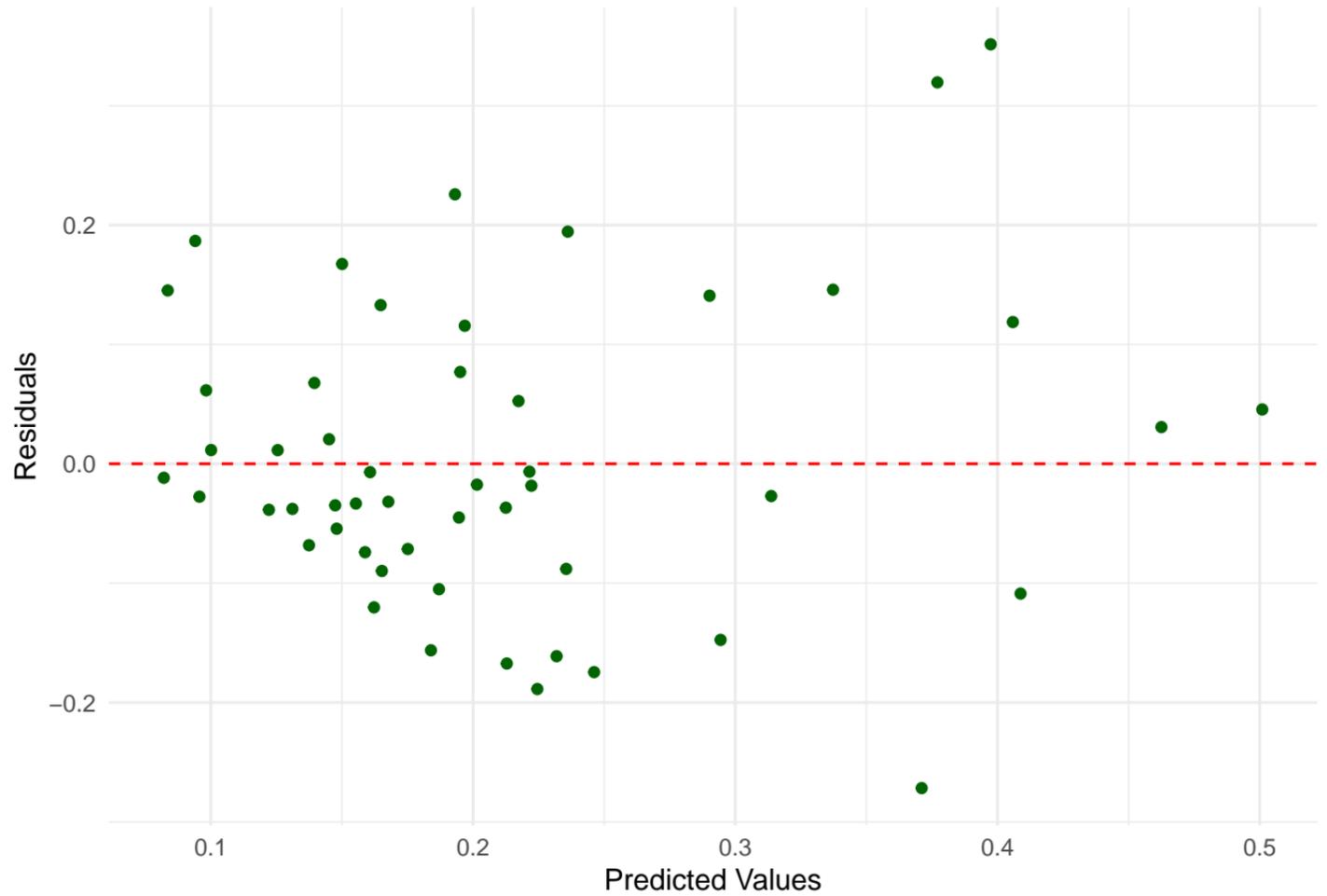

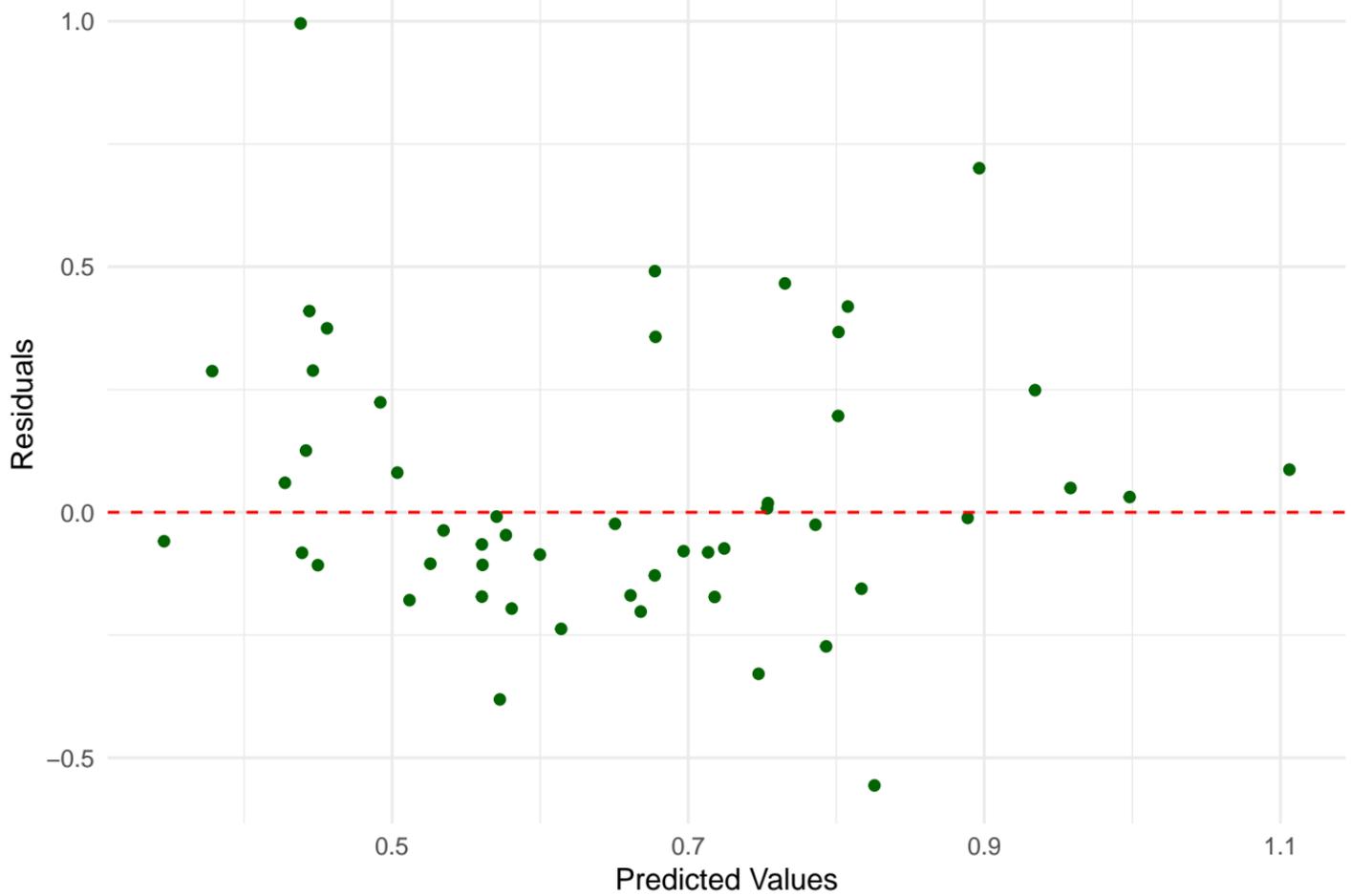

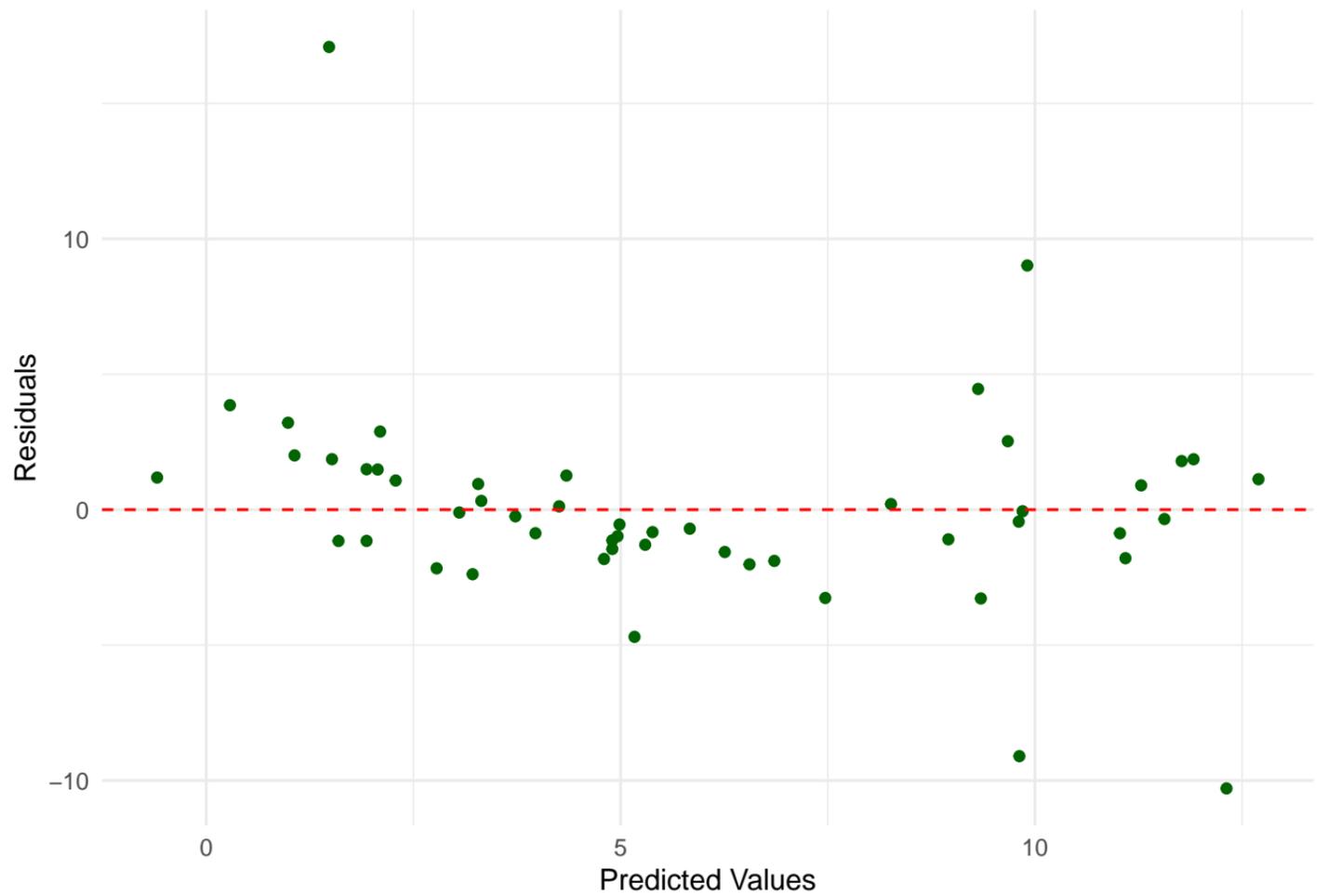

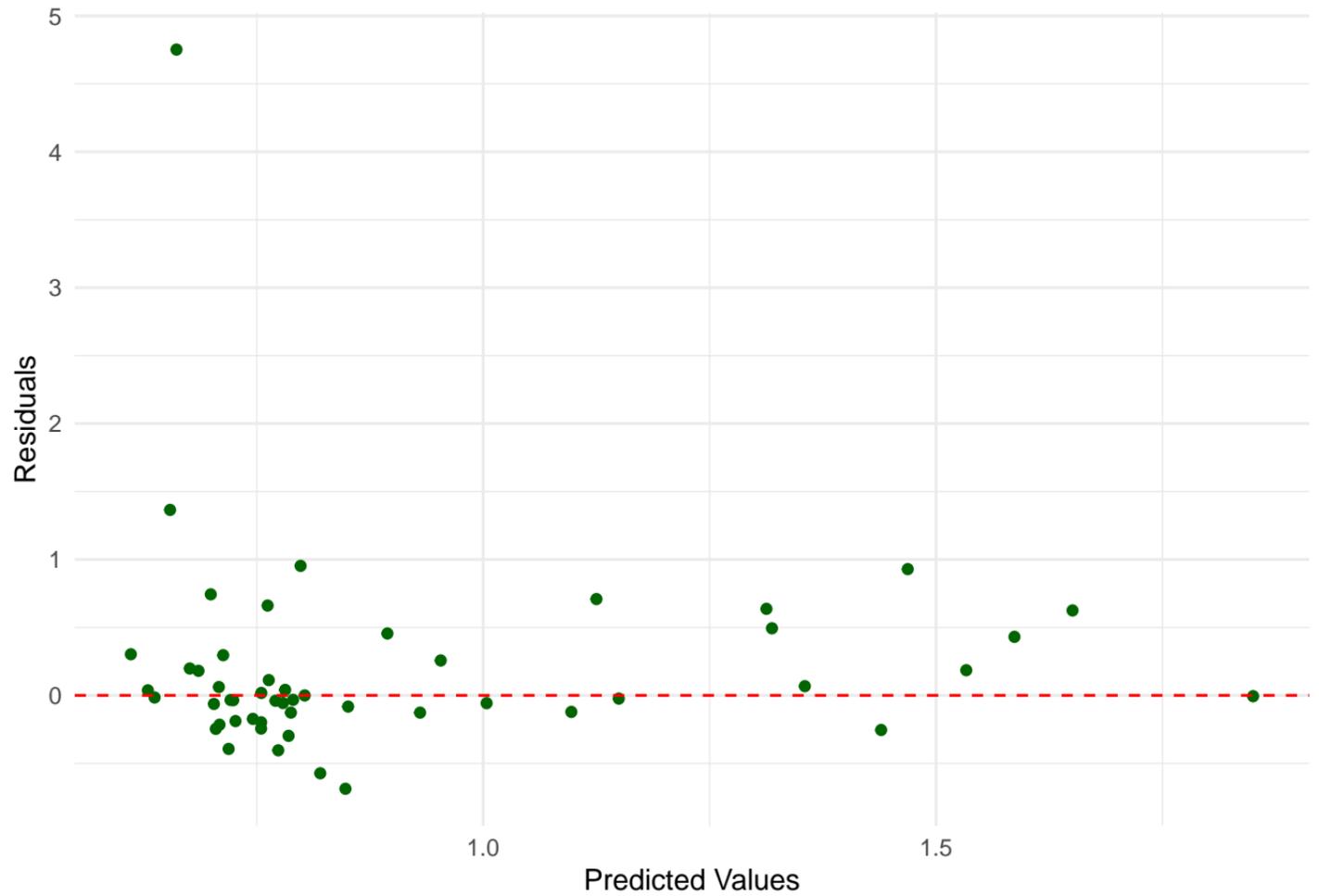

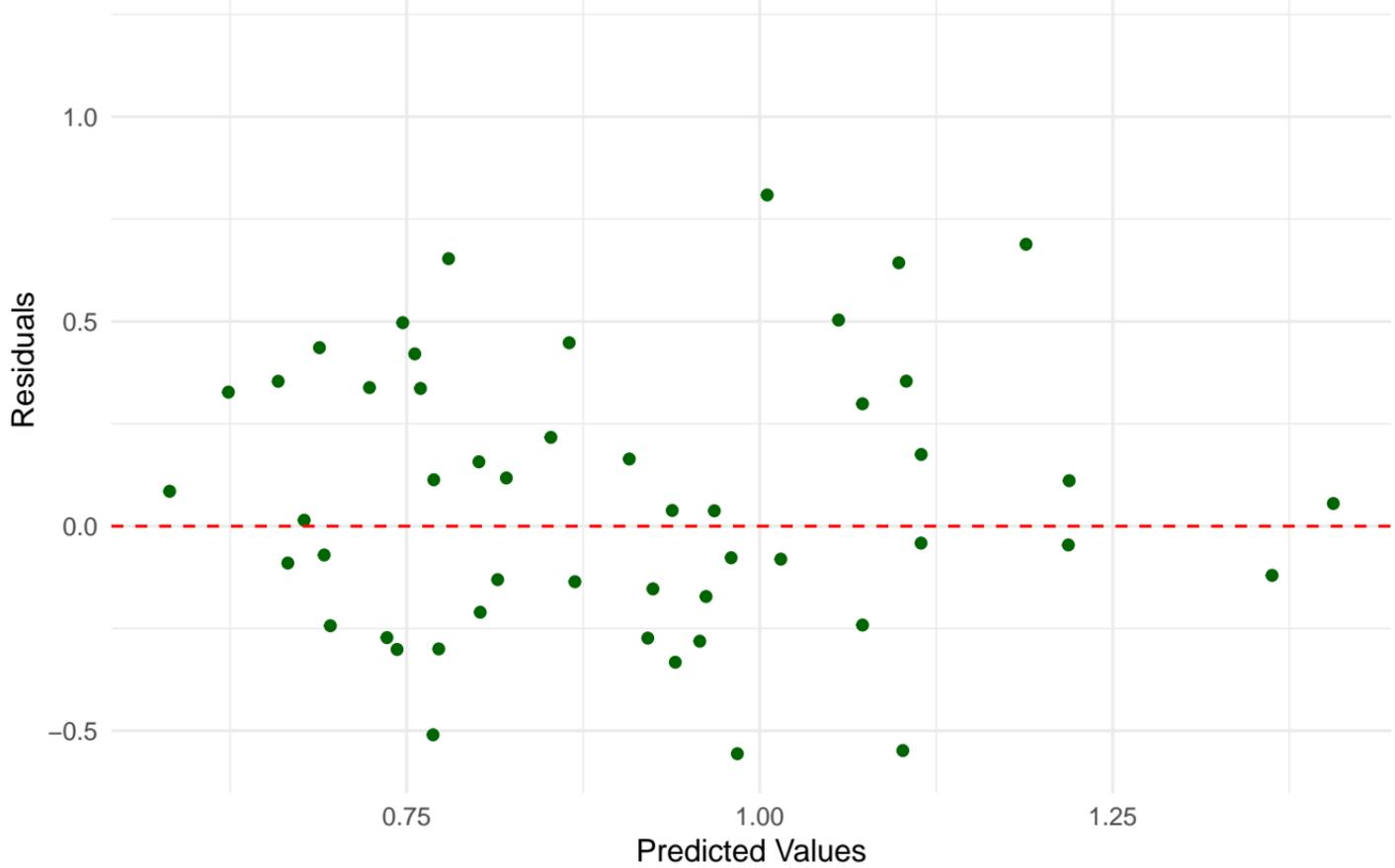

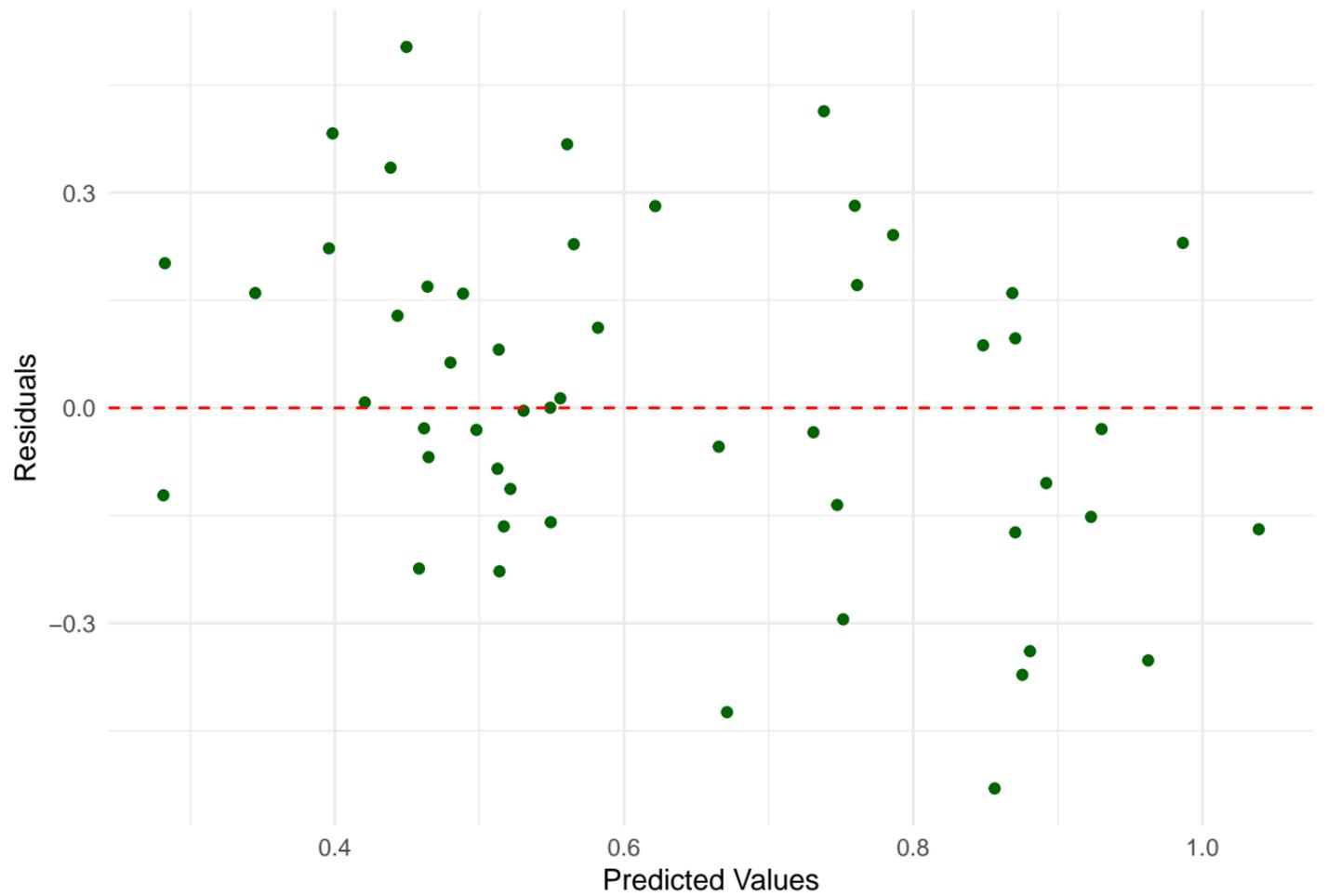

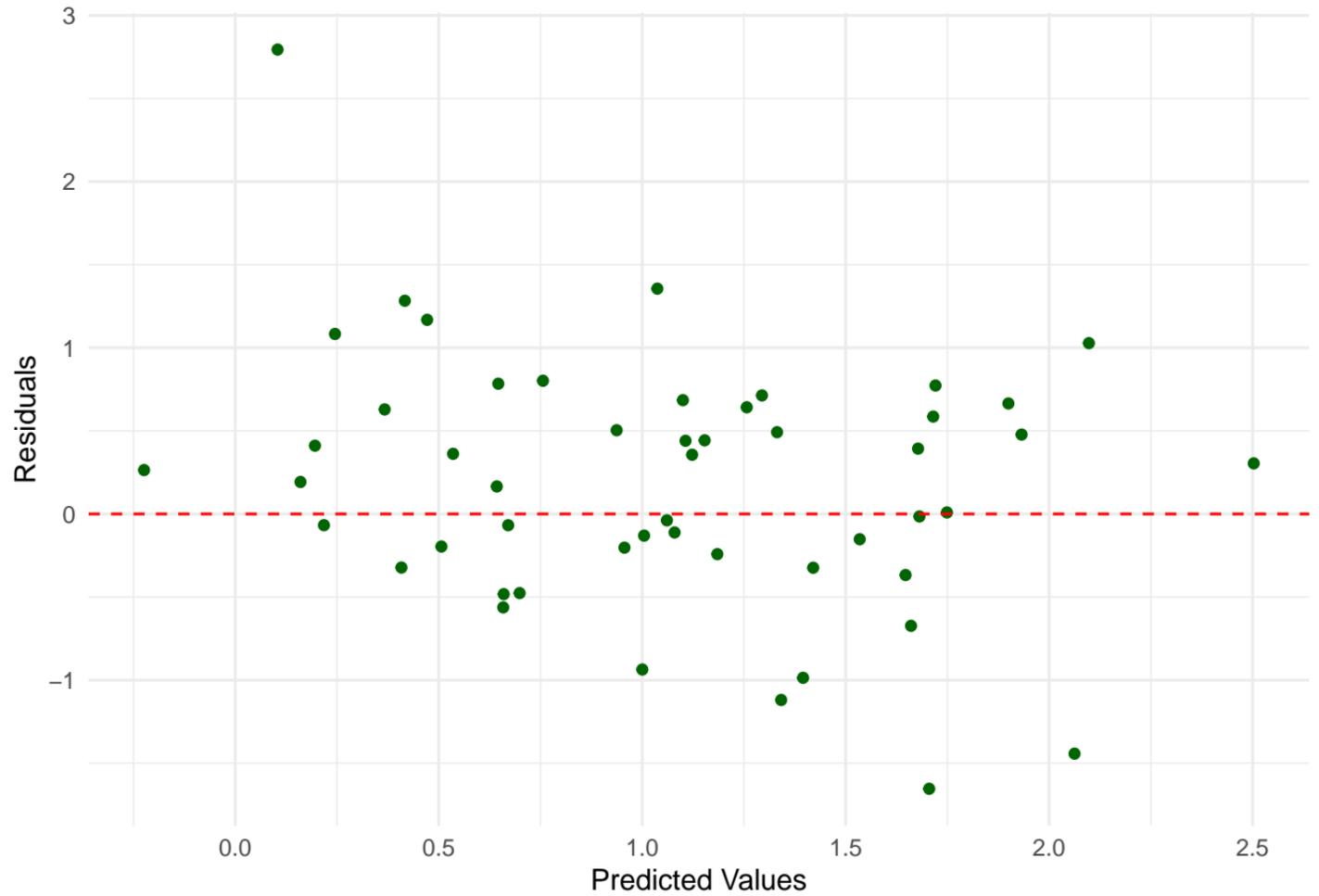

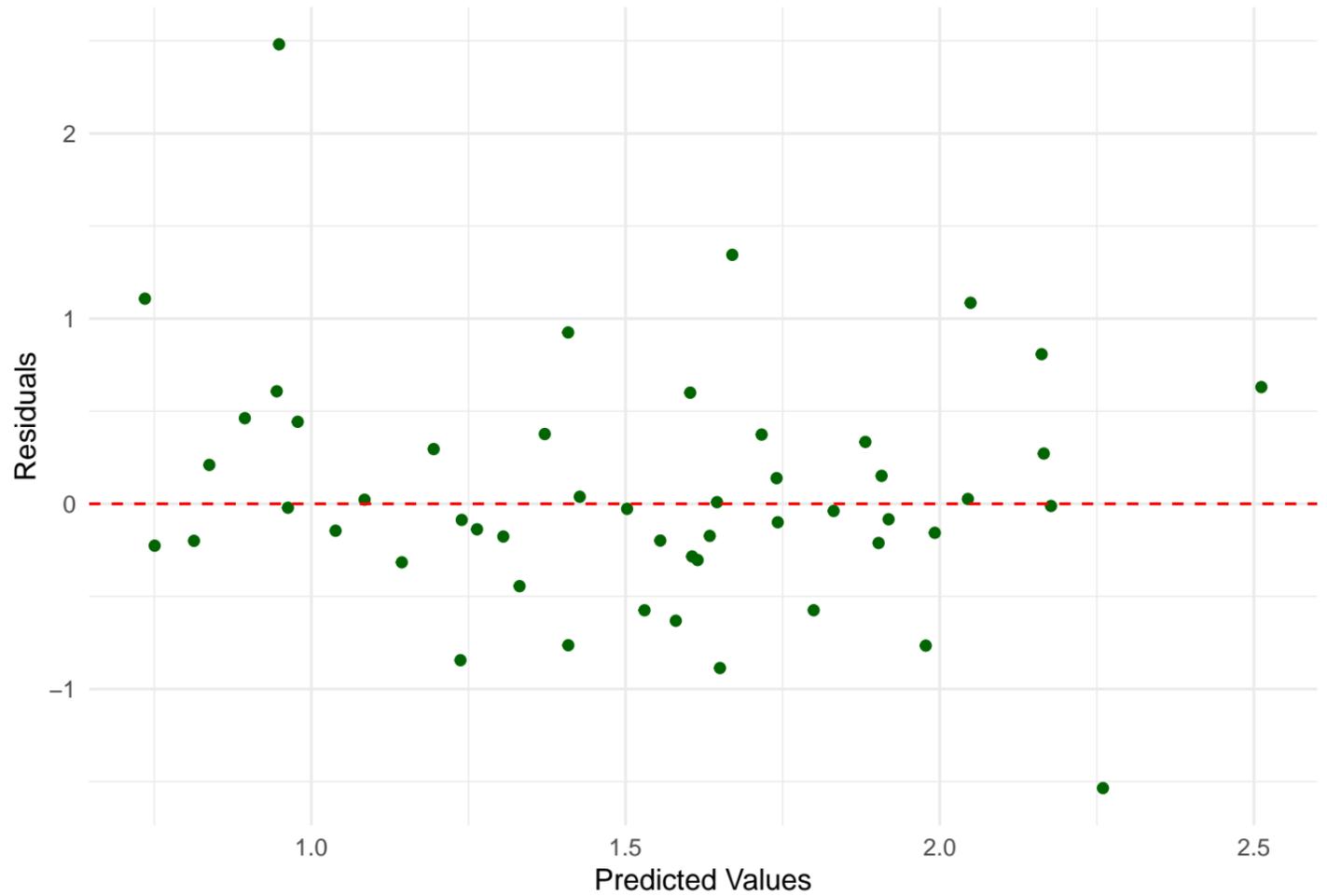

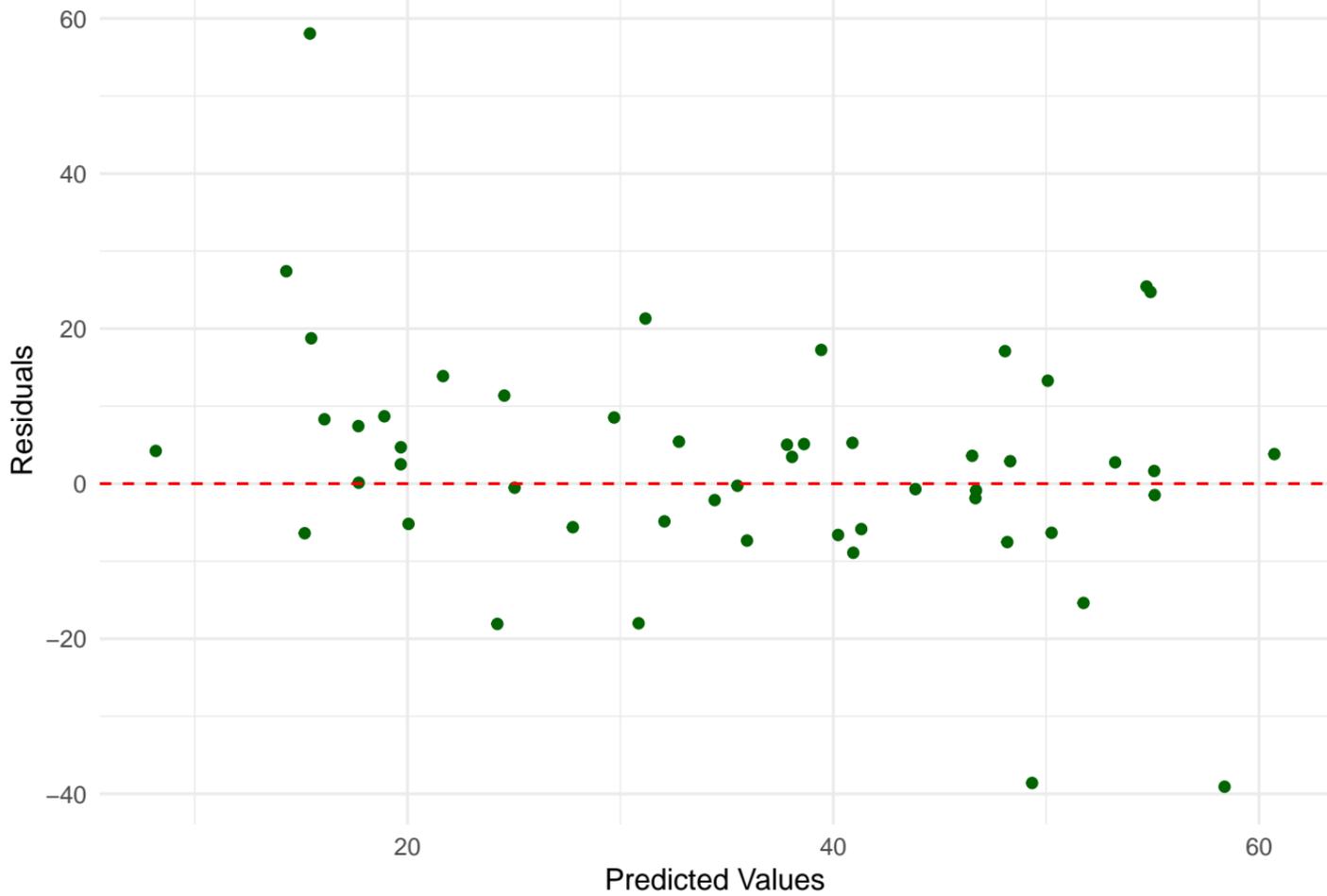